\documentclass{article}
\usepackage{amssymb,amsmath}
\usepackage{authblk}
\usepackage{graphicx}
\usepackage[utf8]{inputenc}
\usepackage[english]{babel}
\usepackage{natbib}
\usepackage{booktabs,multirow}
\usepackage{subcaption}
\usepackage{adjustbox}
\usepackage{multirow}
\usepackage{multicol}

\usepackage{tikz}
\usetikzlibrary{shapes,arrows}

\usepackage{geometry}
\usepackage{lineno}

\title{Bacteriophage effect on parasitism resistance}
\date{\today}

\author[1]{Gabriel R. Palma}
\author[3, 4]{Renato M. Coutinho}
\author[2]{Wesley A. C. Godoy}
\author[2]{Fernando L. Cônsoli}
\author[3, 4]{Roberto A. Kraenkel}
\affil[1]{
    Hamilton institute\\
    Maynooth University, Maynooth, Ireland}
\affil[2]{
    Departamento de Entomologia e Acarologia\\
    Universidade de S\~{a}o Paulo, Piracicaba, Brazil}
\affil[3]{
    Centro de Matem\'atica, Computa\c{c}\~ao e Cogni\c{c}\~ao (CMCC),
    Universidade Federal do ABC\\
    Av. dos Estados, 5001, 09210-580 Santo Andr\'e, Brazil
}
\affil[4]{
    Instituto de Física Teórica\\
    Universidade Estadual Paulista, São Paulo, Brazil
}

\begin{document}

\maketitle

\begin{abstract}
Many studies have shown that the protection of the host \textit{Acyrthosiphon pisum} (Hemiptera, Aphididae) against the parasitoid \textit{Aphidius ervi} (Hymenoptera, Braconidae) is conferred by the interaction between the secondary endosymbiont \textit{Hamiltonella defensa} and the bacteriophage \textit{APSE} (\textit{Acyrthosiphon pisum} secondary endosymbiont). This interaction consists of the production of toxins by the endosymbiont's molecular machinery, which is encoded by the inserted \textit{APSE} genes. The toxins prevent the development of the parasitoid's egg, conferring protection for the host. However, the effects of this microscopic interaction on host-parasitoid dynamics are still an open question. We presented a new mathematical model based on the bacteriophage effect on parasitism resistance. We identified that the vertical transmission of the bacteriophage and the host survival after the parasitoid attack are potential drivers of coexistence. Also, we showed that the vertical transmission of \textit{H. defensa} is proportional to the time that the protected population became extinct. Our results showed that the protected and unprotected hosts' survival after the parasitoid attack is fundamental to understanding the equilibrium of long host-parasitoid dynamics. Finally, we illustrated our model considering its parameters based on experiments performed with \textit{A. pisum} biotypes \textit{Genista tinctoria} and \textit{Medicago sativa}. 

\end{abstract}

\modulolinenumbers[5]

\section{Introduction}
Symbiosis is a type of interaction that mutual benefits between two or more individuals who live close together \citep{Moran2006,bourtzis2008,Haine2008, Martin2012}. It can be endosymbiotic when one of the individuals (symbiont) lives inside the body of the other individual (host), either inside their cells or extracellularly \citep{bourtzis2008}. The interaction between symbiont and host may be mandatory when one or both individuals depend on each other for survival or reproduction or optional when the interaction is not strictly necessary \citep{Moran2006, bourtzis2008, Hosokawa2020}. A mandatory endosymbiont will, in most cases, be present in all individuals of the host population since hosts who do not have the symbiont will not be able to survive \citep{Hosokawa2020}.

In insects, a well-known case of mandatory endosymbiosis is the interaction between aphids and the endosymbiotic bacterium \textit{Buchnera aphidicola} \citep{Gil2019}. The bacterium is strictly necessary for the production of amino acids that the aphid is unable to produce while benefiting from the resources used and processed by the aphid, characterizing the process with mutual benefits through interaction \citep{Gil2019}. On the other hand, the result of this interaction can also imply antagonistic actions, depending essentially on how the benefits are obtained \citep{Moran2006, Haine2008, Hosokawa2020}.

During these interactions, the occurrence and permanence of symbionts in a host population are strongly dependent on how the endosymbionts are transmitted between individuals in the host population \citep{Vorburger2017, Russell2019, Hosokawa2020}. The transmission can be vertical when passed from mother to progeny, or horizontal, through direct contact between individuals, mediated by secretion or excrement and transmitted by parasitoids \citep{Haine2008, Gehrer2012, Kaech2021}. The transmission can also be a combination of these processes \cite{Haine2008, Gehrer2012, Kaech2021}. For optional vertically transmitted endosymbiont, the optional nature of the organism itself should influence your transmission rate \citep{bourtzis2008}. Depending on the rate of transmission, the endosymbiont population may even be locally extinct in case the host is unable to pass it on to the next generations \citep{bourtzis2008,kwiatkowski2012}. The study of mechanisms capable of influencing the success or extinction of endosymbionts in insects comprises an active area of research aimed at investigating the interactive dynamics of organisms \citep{Haine2008,brownlie2009}.

The dynamics of the interaction between a given host and its endosymbionts can also be important for understanding other trophic relationships in the community. A good example, particularly relevant for a scenario involving insects, is the parasitoid-symbiont-host system \citep{kwiatkowski2012,vorburger2013}. This system has already been investigated considering the host \textit{Acyrthosiphon pisum} (Hemiptera, Aphididae), the parasitoid \textit{Aphidius ervi} (Hymenoptera, Braconidae) and the secondary endosymbiont \textit{Hamiltonella defensa}, a gamma-proteobacteria, \citep{kwiatkowski2012, Kaech2021RNA}. Many results have shown that the interaction between this endosymbiont and the bacteriophage named \textit{APSE} (\textit{Acyrthosiphon pisum} secondary endosymbiont) impacts the host protection against parasitoids \citep{oliver2009, Kaech2021RNA}. This interaction consists of the production of toxins by the endosymbiont’s molecular machinery, which is encoded by the inserted \textit{APSE} genes. The toxins prevent the development of the parasitoid’s egg, conferring protection for the host \citep{Haine2008, Leybourne2020, Kaech2021RNA}.



In the example mentioned, the protection that the host \textit{A. pisum} receives only occurs when the endosymbiont \textit{H. defensa} is infected with the bacteriophage, \textit{APSE} (\textit{Acyrthosiphon pisum} endosymbiont secondary) \citep{VanDerWilk1999, Rajarajan2011}. Thus, the presence of the bacteriophage is fundamental for encoding the gene that produces the toxins that protect the hosts \citep{oliver2009, Society2017, Oliver2019}. The bacteriophage belongs to the Podoviridae family \citep{VanDerWilk1999, Rajarajan2011} and different varieties of this bacteriophage confer different levels of protection given the type of toxin that is produced. The endosymbiotic molecular machinery, once infected by the variety \textit{APSE}-2, produces a homologue of the cytolethal distention toxin (cdtB), causing the host protection of approximately $40\%$ against the parasitoid attack. Whereas the variety \textit{APSE}-3 encodes a YD-repeat that has toxic proteins, and when the symbiosis occurs between the host and the infected endosymbiont, the protection against the parasitoid attack is approximately $85\%$ \citep{oliver2009}. This protection variation is also found within different hosts biotypes \citep{Sochard2020}.



This microbial interaction causes directly or indirectly interference in interspecific interactions within the community. In general, organisms can modify the use of host plants by phytophagous insects, provide resistance to natural enemies, and also reduce global genetic diversity or gene flow between populations within some species \citep{ferrari2011,Frago2017}. These actions can occur through changes between the sex ratio of insects, caused, for example, by the bacteria of genera \textit{Wolbachia}, or even by the history of genetic relationship between the host and the symbiont \citep{Moran2006, bourtzis2008, ferrari2011}. Actions of this nature are often highly relevant to the permanence of symbionts in their hosts. The implications of the symbiont's permanence within the system, as well as their effect on the parasitoid-host relationship \citep{kwiatkowski2012}, give the system significant complexity.

The complexity encountered at the system involving the aphid \textit{Acyrthosiphon pisum}, the parasitoid \textit{Aphidius ervi}, the endosymbiont \textit{H. defensa} and the bacteriophage \textit{APSE} have a direct impact on biological controls methods\citep{Vorburger2018}. Given that releasing parasitoids at agroecosystems is a common procedure \citep{Vorburger2018, Giunti2015, Parra2019, Leung2020} to reduce pest population, the host protection against the parasitoids makes a significant impact on pest control\citep{Godfray1994, Giunti2015, Leung2020}. Thus, understanding the host protection evolution is fundamental to optimising biological control methods and reducing economic damage caused by these pests. Also, this dependent protection provided by the microbial interactions turns pest control into a challenging research question.



The whole scenario involving the simultaneous occurrence of protected and unprotected hosts requires careful analysis for the correct interpretation of the processes and phenomena involved, lacking the use of analytical tools with the potential to describe the complex mechanisms involved in the system \citep{Leung2020}. It can be observed in several scales \citep{Rocha2018} and to understand this system. It is necessary to use appropriate tools to identify and interpret ecological patterns more effectively. Mathematical and computational models are commonly worn among the analytical tools available to investigate systems of this nature \citep{Ferreira2014}. The use of mathematical models to describe ecological patterns in dynamical systems has brought a significant scientific contribution due to the flexibility of these tools to develop algorithms capable of analyzing different ecological phenomena in time and space. Mathematical models can be used to describe ecological processes and predict population trends. The use of this resource has been increasing in recent decades in response to the growing demand for the formalization of population processes, with the possibility of simulating biological scenarios. The understanding of fundamental population aspects and community functions using ecological mathematical formalism has produced actual results to compose ingredients of ecological theory, capable of covering a broad spectrum of issues, ranging from ecological space and time patterns of populations and communities to epidemiological aspects of trophic networks \citep{Ferreira2014}.

The perspective of mathematical modelling brings the idea of reorganizing dynamic systems considering dimensions and scales that can be investigated \citep{Ferreira2014}. This new condition under which systems can be visualized allows new interpretations in a gradual and oriented way so that the population variations of systemic members can be understood as coming from endogenous and exogenous forces governing the system as a whole \citep{Ferreira2014}. The expectation of understanding the influence of the endosymbiont on the host-parasitoid system ecologically can be met through the use of mathematical models \citep{kwiatkowski2012}(Jones and Boots, 2007; Kwiatkowski and Vorburger, 2012). Issues of theoretical and applied relevance can be investigated from models capable of analyzing the population dynamics of the system, including the influence of the vertical transmission of endosymbionts. Here we aim to study the influence of the bacteriophage \textit{APSE} on the host-symbiont-parasitoid system and its impact on the protection evolution. Thus, we introduced a new mathematical model that contains the microbial interaction responsible for conferring protection to the host \textit{A. sifum} against the parasitoid \textit{A. ervi}.

\section{Methods}

\subsection*{Model formulation}

To model the system involving the aphid \textit{Acyrthosiphon pisum} and the microorganisms responsible for its protection against the parasitoid \textit{Aphidius ervi} \cite{oliver2009} we considered the results provided by the \cite{oliver2009} and added the microorganisms interaction by including the bacteriophage \textit{APSE} and the secondary symbiont \textit{Hamiltonela defensa} on the host-parasitoid system. To consider these elements, the model consists of three differential equations representing the host population of \textit{Acyrthosiphon pisum} which is infected with the endosymbiont \textit{Hamiltonela defensa} ($H$), population of \textit{Acyrthosiphon pisum} which is infected with the endosymbiont \textit{Hamiltonela defensa} plus the bacteriophage \textit{APSE} ($V$) and the non-infected population of \textit{Acyrthosiphon pisum} ($S$). The model is composed of the following system:

\begin{equation}
\begin{aligned}    
\frac{dS}{dt} &= b_S S\left(1-\frac{N}{K}\right)\exp{\frac{\log{\left(\alpha\right)} P}{\beta N + P}} - \mu S\\  
&+ (1 - \tau_H) b_H H \left(1- \frac{N}{K}\right) \exp{\frac{\log{\left(\alpha\right)} P}{\beta N + P}}\\ 
&+ (1-\tau_V)b_V V \left(1-\frac{N}{K}\right)\exp{\frac{\log{\left(\alpha\right)} P}{\beta N + P}}~, 
\\
\frac{dH}{dt} &= \tau_H b_H H \left(1- \frac{N}{K}\right) \exp{\frac{\log{\left(\alpha\right)} P}{\beta N + P}} - \mu H
\\
&+ (1-\tau_X)\tau_V b_V V \left(1-\frac{N}{K}\right)\exp{\frac{\log{\left(\alpha\right)} P}{\beta N + P}}~,
\\
\frac{dV}{dt} &=\tau_X \tau_V b_V V \left(1-\frac{N}{K}\right)\exp{\frac{\log{\left(\theta\right)} P}{\beta N + P}} - \mu V~.
\end{aligned}
\end{equation}
We considered the logistic growth $b_S\left( 1- \frac{N}{K}\right)$, where $b_s$, $b_h$ and $b_v$ are growth rates, $N = S + H +V$ is the total of hosts and $K$ is the carrying capacity for the system. We consider the logistic growth based on the fact that the population density of hosts depends on the presence of resources and the space restriction encountered during its life cycle. To indicate the protection against parasitoids, we included the survival rate as $\exp\frac{-\alpha P}{\beta N+P}$, where $P$ is the number of parasitoids, $\alpha$ is the percentage of unprotected hosts ($S$ and $H$) killed by its attack, and $\beta$ is the oviposition ratio. For the hosts that carry both \textit{APSE} plus \textit{Hamiltonela defensa}, we used the parameter $\theta$ representing the percentage of protected hosts ($V$) killed by parasitoids. To represent the protection of this population, we set $\theta \geq \alpha$.

\begin{figure}[htb]
\centering
\tikzstyle{block} = [rectangle, draw, fill=blue!20, text width=9em, text
    centered, rounded corners, minimum height=3em]
\tikzstyle{line} = [draw, -latex']
\tikzstyle{dline} = [draw, -latex-']
\tikzstyle{cloud} = [rectangle, draw, fill=red!20, text width=9em, text
    centered, node distance=3cm, minimum height=3em]

\begin{tikzpicture}[node distance = 2cm, auto]
    \node [block] (S) {Adult host with no \emph{H. defensa} ($S$)};
    \node [block, right of=S, node distance=5cm] (H) {Adult host with
        \emph{H. defensa} but not \emph{APSE} ($H$)};
    \node [block, right of=H, node distance=5cm] (V) {Adult host with both
        \emph{H. defensa} and \emph{APSE} ($V$)};
    \node [cloud, below of=S] (LS) {Nymphs with no \emph{H. defensa}};
    \node [cloud, below of=H] (LH) {Nymphs with
        \emph{H. defensa} but not \emph{APSE}};
    \node [cloud, below of=V] (LV) {Nymphs with both
        \emph{H. defensa} and \emph{APSE}};
    \node [cloud, below of=LH] (P) {Parasitoid};
    \path [line] (S) -- node {$b_S$} (LS);
    \path [line] (H) -- node [near end] {$b_H \tau_H$} (LH);
    \path [line] (V) -- node {$b_V \tau_V \tau_X$} (LV);
    \path [line] (LS) to [out=120,in=-120] (S);
    \path [line] (LH) to [out=120,in=-120] (H);
    \path [line] (LV) to [out=120,in=-120] (V);
    \path [line] (H) -- node [pos=0.6,xshift=-1.4cm, yshift=0.6cm] {$b_H(1-\tau_H)$} (LS);
    \path [line] (V) -- node [pos=0.8,xshift=-0.3cm,yshift=0.3cm] {$b_V(1-\tau_V)$} (LS);
    \path [line] (V) -- node [pos=0.7,xshift=-0.2cm] {$b_V \tau_V (1-\tau_X) $} (LH);
    \path [line,dashed] (P.west) -| node [near end] {$\alpha$} (LS);
    \path [line,dashed] (P.north) -| node [near end] {$\alpha$} (LH);
    \path [line,dashed] (P.east) -| node [near end] {$\theta$} (LV);
    \node [above of=S] (mS) {};
    \path [line,dashed] (S.north) to node {$\mu$} (mS);
    \node [above of=H] (mH) {};
    \path [line,dashed] (H.north) to node {$\mu$} (mH);
    \node [above of=V] (mV) {};
    \path [line,dashed] (V.north) to node {$\mu$} (mV);
\end{tikzpicture}
\caption{Diagram representing the ingredients of the model. Only the blue boxes are dynamic variables. Full lines represent reproduction and maturation, dotted ones represent mortality.}
\label{fig:model-diagram}
\end{figure}
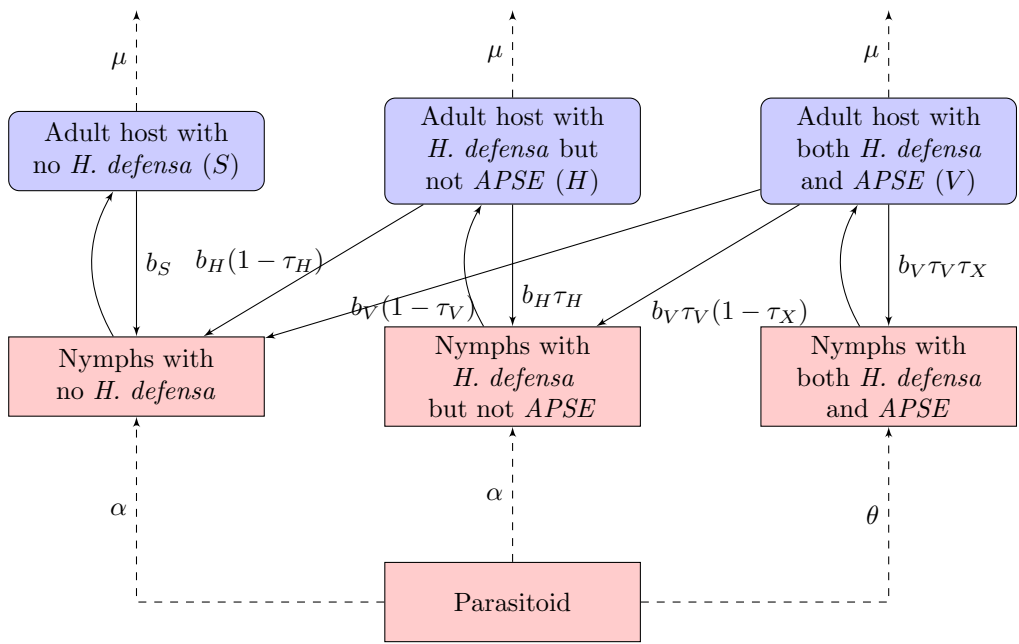

The vertical transmission of \textit{Hamiltonela defensa} can occur in two forms by our model. The first considers that the secondary endosymbiont can be transmitted to an offspring by adults that has only \textit{Hamiltonela defensa} ($\tau_H$) or by adults that has both \textit{Hamiltonela defensa} plus the bacteriophage \textit{APSE} ($\tau_V$). We also considered the vertical transmission of the bacteriophage \textit{APSE} by the parameter $\tau_X$. These vertical transmissions are based on the anatomy of \textit{Acyrthosiphon pisum} that indicates these possible types of transmissions. More specifically, we are considering that during the process of nymphs birth, the aphid embryo can touch or not the mycetocytes (Cells carrying secondary endosymbiont and the bacteriophage) to receive these microorganisms \citep{VanDerWilk1999}. These transmissions can be clarified by the figure~\ref{fig:model-diagram}, where we represent all possibilities of microorganisms' vertical transmissions and other components of our model. Finally, we considered that the natural mortality of each population of hosts is the same $\mu$ based on the fact that the main difference between these populations only occurs with parasitoids.





\subsection*{Parameter setting}
The growth rate of each population $b_s$,$b_h$ and $b_v$ are here determined by \textit{Acyrthosiphon pisum} biology and the constitutive and induced cost of the harbouring the endosymbiont \citep{Hutchison1985, Lu2008, weldon2013phage, Kaech2022}. Thus, we set $b_s = 0.4$, $b_v = 0.3$ and $b_h = 0.2$ considering a constitutive cost of $c_c=\frac{b_s - b_h}{b_s} = 0.5$ and a induced cost $c_i = \frac{b_s - b_v}{b_s} = 0.25$. We also set the carrying capacity $k=15,000$ \citep{kwiatkowski2012} and the death rate $\mu = 0.05$ \citep{Hutchison1985, Lu2008}. The vertical transmission rates $0 \leq \tau_H \leq 1$, $0 \leq \tau_V \leq 1$, $0 \leq \tau_X \leq 1$ are positive parameters. Here we set $\tau_H=0.995$ based on the feasibility of vertical transmission of \textit{H. defensa} \citep{Vorburger2017}. We also considered $\tau_V \leq \tau_H $ based on the fact that the bacteriophage can kill some symbionts and create difficulties during the vertical transmission of the endosymbiont. Considering the vertical transmission, the anatomy of the hosts suggests that these vertical transmission parameters are closed, thus we set $\tau_V=0.95$ and $\tau_X=0.995$. However we could not found it in literature and we selected this parameter empirically based on the insect biology. To overcome this problem, we analyzed the parameters space of $\tau_X$ and $\tau_V$ by bifurcation analysis.

We considered two biotypes of the host \textit{Acyrthosiphon pisum} for the parameter choice. The first one is the \textit{Ganista tinctoria} biotype presented in the table 1. This biotype has the secondary endosymbiont fixed into their natural population \citep{Sochard2020}, thus we set the model variables $S=0$, $H=500$ and $V=2500$. To select the host survival rate after parasitoid attack we used the experiments from \cite{Sochard2020} to obtain $\alpha=0.549$ and $\theta=0.676$, more specifically, we used the data from \cite{sochard_data_2020} \textit{G. tinctoria} biotype. For this task we selected hosts infected with \textit{H. defensa} plus \textit{APSE} ("H-Ms2" in the dataset) and uninfected biotype ("Cured" in the dataset) \textit{Ganista tinctoria} (In the data set they are named "G.tinctoria"). During the experiment the hosts were artificially infected with \textit{H. defensa} strains from another aphid biotype named \textit{Medicago sativa} (For more details see \cite{Sochard2020}). Using $N$ experiments, which briefly consists on offering $M=15$ aphids to \textit{A. ervi} female in a glass Petri dish containing a leaf disk of \textit{V. faba}, we used the number of emerged parasitoids from uninfected hosts $P^s$ to compute $\alpha=-\log{\left(\frac{\sum_{n=1}^{N}\left( M-P^{s}_n\right)}{N}
\right)},$ and the number of emerged parasitoids from infected hosts $P^{v}$ to compute $\theta=-\log{\left(\frac{\sum_{n=1}^{N}\left( M-P^{v}_n\right)}{N}\right)
}.$

The second one is the \textit{Medicago sativa} biotype presented in table 1. In natural populations \textit{H. defensa} has intermediate to high frequencies in this biotype, thus we set the model variables $S=500$, $H=500$ and $V=2000$. The survival rates were obtained using the results provided by the first experiment of the work of \cite{Dion2011}, more specifically, the results encountered in the section \textit{Experiment 1: Aphid clone resistance measurement}. Using the equation $2$ coupled with the results of $P^{s}=143$ parasitoids emerged from $M=200$ uninfected hosts we obtained $\alpha=0.285$. We used the equation $3$ and the results of $P^{v} = 33$ parasitoids emerged from $M=400$ to obtain $\theta=0.917$. Finally, for both biotypes, we set $\beta=\frac{1}{105}$ and performed a bifurcation analysis to observe the influence of these parameters on the stability of the system.

\begin{table}[ht]
\caption{Description of parameters and variables used for the host-virus-parasitoid model. The values of each parameter are presented for the species \textit{Medicago sativa} and \textit{Genista tinctoria} hosts biotype. These values were obtain based on the dataset available in the work of \cite{Sochard2020} and \cite{Dion2011}.}
\label{parameters}
\centering
\begin{adjustbox}{max width=\textwidth}
\begin{tabular}{llrr}
\hline
Symbol & 
Meaning & 
\multicolumn{2}{c}{Values} 
\\
\hline
Model variables: & & &
\\
\hline
& & \textit{Genista tinctoria}& \textit{Medicago sativa}
\\
$S$ & Uninfected hosts & 0 (Initial) & 500 (Initial)
\\

$H$ & Hosts infected with \textit{Hamiltonela defensa} & 500 (Initial)&500 (Initial)
\\

$V$ & Hosts infected with \textit{Hamiltonela defensa} and \textit{APSE} & 2500 (Initial) & 2500 (Initial)
\\

Parameters:
\\
\hline
$bs$ & Growth rate of population $S$ & 0.4 & 0.4
\\
$bh$ & Growth rate of population $H$ & 0.2 & 0.2
\\
$bv$ &  Growth rate of population $V$ & 0.3 & 0.3
\\ 
$\mu$ & Mortality rate & 0.05 & 0.05
\\
$\tau_H$ & Vertical transmission of \textit{H. defensa} by $S$ & 0.995 & 0.995
\\
$\tau_V$ & Vertical transmission of \textit{H. defensa} by $V$ & 0.950 & 0.950
\\
$\tau_X$ & Vertical transmission of \textit{APSE} & 0.995 & 0.995
\\
$K$ & Carrying capacity & 15.000 & 15.000
\\
$P$ & Parasitoid population & 200 & 200
\\
$\alpha$ & Hosts survival after parasitoid attack & $0.549$ & $0.285$
\\
$\theta$ & Hosts survival after parasitoid attack & $0.676$ & $0.917$
\\
$\beta$ & Oviposition ratio & $\frac{1}{105}$ & $\frac{1}{105}$
\\\\
\hline
\end{tabular}
\end{adjustbox}
\end{table}

\subsection*{Analyses}
\subsubsection*{Bifurcation}

To perform the bifurcation analysis we used the parameters obtained from both host biotypes \textit{Genista tinctoria} and \textit{Medicago sativa}. For that, we selected each parameter to analyse its effect on the dynamics stability in both scenarios. Using the parameter and variables from table 1, we analyse the following parameters spaces $P=\left( 1, 2, \ldots, 100\right)$, $\mu =\left( 0, 0.01, \ldots, 1\right)$, $\tau_H=\left( 0, 0.01, \ldots, 1\right )$, $\tau_V=\left( 0, 0.01, \ldots, 1\right )$, $\tau_X=\left( 0, 0.01, \ldots, 1\right)$, $\alpha=\left( 0, 0.01, \ldots, 1\right)$, $\theta=\left( 0, 0.01, \ldots, 1\right )$ and $\beta=\left( 0, 0.01, \ldots, 1\right)$. For each parameter we computed $5000$ model iterations and selected the number of individuals for $S$, $V$ and $H$ in the last iteration (The $5000^{\mbox{th}}$ iteration). We repeated this process for each value of the the selected parameter space and we also performed this process using the parameters and variable from table 1. 

Finally, we performed two dimensional bifurcation analysis with the parameters $\left( P, \tau_V\right)$, $\left( P, \tau_X\right)$, $\left( \tau_V, \tau_X\right)$, $\left( \alpha, \theta\right)$ and $\left( c_c, c_1\right)$. This briefly consists on computing $5000$ iterations to obtain $S$, $V$ and $H$. The difference here is that we used a grid $P=\left(0, 1, \ldots, 1000 \right)$, $\alpha=\left(0, .01, \ldots, 1 \right)$, $\theta=\left(0, .01, \ldots, 1 \right)$, $c_c=\left(0, .01, \ldots, 1 \right)$, $c_i=\left(0, .01, \ldots, 1 \right)$ and $\left(0, 0.01, \ldots, 1 \right)$ for $\tau$ parameters. Thus, for each pair of values we obtained $S$, $V$ and $H$ in the last iteration to evaluate the model stability. After that, we transform the density of all populations in a binary vector that represents the the presence of all populations ($\boldsymbol{v}^{'}=\left[1, 1, 1\right]$), only $S$ ($\boldsymbol{v}^{'}=\left[1, 0, 0\right]$), only $H$ ($\boldsymbol{v}^{'}=\left[0, 1, 0\right]$), only $V$ ($\boldsymbol{v}^{'}=\left[0, 0, 1\right]$), $S$ with $H$ ($\boldsymbol{v}^{'}=\left[1, 1, 0\right]$), $S$ with $V$ ($\boldsymbol{v}^{'}=\left[1, 0, 1\right]$) and $H$ with $V$ ($\boldsymbol{v}^{'}=\left[0, 1, 1\right]$). Using these vectors we could represent the presence and absence of the host populations for the grid of parameters. We repeat this process for both host biotypes fixing the other parameters of the model based on table 1 as previously.

\subsubsection*{Protection evolution}

To analyse the time required to extinct the hosts infected with \textit{H. defensa} plus \textit{APSE} ($V$) from the system with the absence of parasitoids, we obtained the initial values of our model based on the equilibrium point considering the presence of parasitoids in an attempt to guarantee the high frequency of protected hosts. For that, we selected the $5000^{th}$ densities of $S$, $H$ and $V$. Then, we used the parameters presented in table 1 to calculate the number of model iterations required for $V$ reaches zero. To better observe the influence of the vertical transmission of \textit{APSE} and the vertical transmission of \textit{H. defensa} from $V$ on the time required to extinct this infected hosts we repeated this process for $\tau_v=\left(0, 0.01, \ldots, 1\right)$ and $\tau_X=\left(0, 0.01, \ldots, 1\right)$. 

\section{Results}

\subsection{Population Dynamics}
Considering both biotypes \textit{Genista tinctoria} and \textit{Medicago sativa} we simulate their dynamics based on the model formulation. Figure 2 showed that the parasitoid influences the model differently based on these biotypes analyzed. For the scenario with $P=200$ parasitoids we observed that the \textit{Genista tinctoria} system had an equilibrium point $\left(S=12000, H=0, V=0 \right)$ and the density of hosts infected with \textit{H. defensa} plus \textit{APSE} initially increase and after, approximately, $600$ model iterations its population reached $0$. On the other hand, the \textit{Medicago sativa} system reached the equilibrium point $\left(S=1900, H=0, V=10000 \right)$ faster than the other biotype showing the importance of the host survival after parasitoid attack $\alpha$ and $\theta$ for our model.

The scenario with absence of parasitoids ($P=0$) showed that both systems tended for the same equilibrium point $\left(S=12500, H=0, V=0 \right)$. The initial $150$ model iterations showed that the \textit{Genista tinctoria} system presented more hosts infected with \textit{H. defensa} plus \textit{APSE} than the other biotype. Also, figure 2 c and d showed that $V$ reached $0$ faster for the \textit{Medicago sativa} as compared with \textit{Genista tinctoria} system. Individually, our results showed that the \textit{Genista tinctoria} system had the same behaviour for both scenarios with the presence or absence of parasitoids. Otherwise, we observed that the presence of parasitoids influenced the persistence of the hosts infected with \textit{H. defensa} plus \textit{APSE} ($V$), given that these hosts reached $0$ faster than the scenario with an absence of parasitoids. On the contrary, the \textit{Medicago sativa} system showed a difference in behaviour with the presence or absence of parasitoids. The hosts infected with \textit{H. defensa} and \textit{APSE} turned out to be more frequent in the system with parasitoids than the other. The equilibrium point $\left(S=1900, H=0, V=10000 \right)$ showed in figure 2 d illustrate this behaviour. 

\begin{figure}[ht]
	\centering
	\begin{subfigure}{.45\textwidth}
		\centering
		\textit{Genista tinctoria}\par\medskip
		\includegraphics[width=1\linewidth]{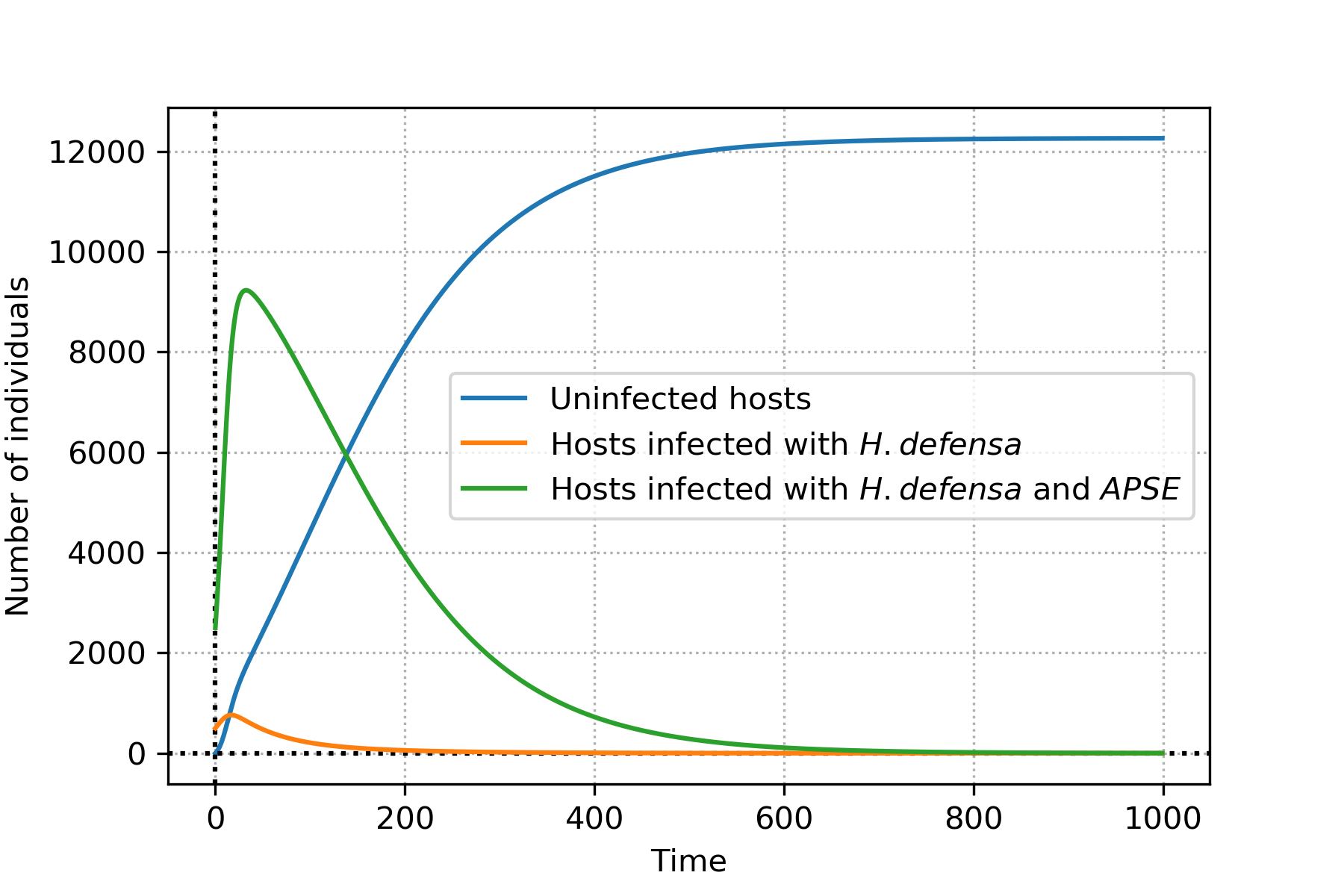}
		
		\caption{ }
	\end{subfigure}
	\begin{subfigure}{.45\textwidth}
		\centering
		\textit{Medicago sativa}\par\medskip
		\includegraphics[width=1\linewidth]{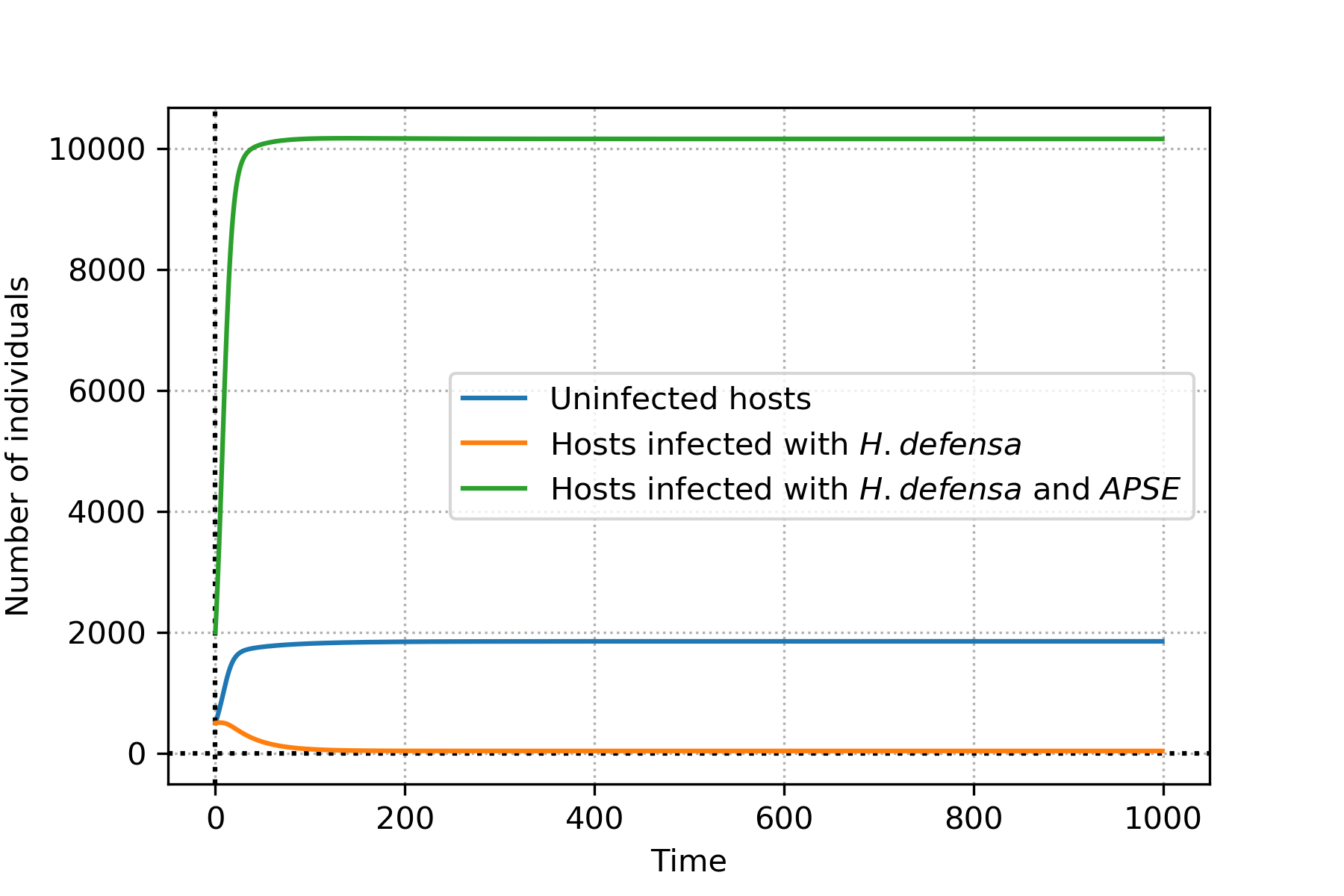}
	
		\caption{ }
	\end{subfigure}
	
	\begin{subfigure}{.45\textwidth}
		\centering
		\includegraphics[width=1\linewidth]{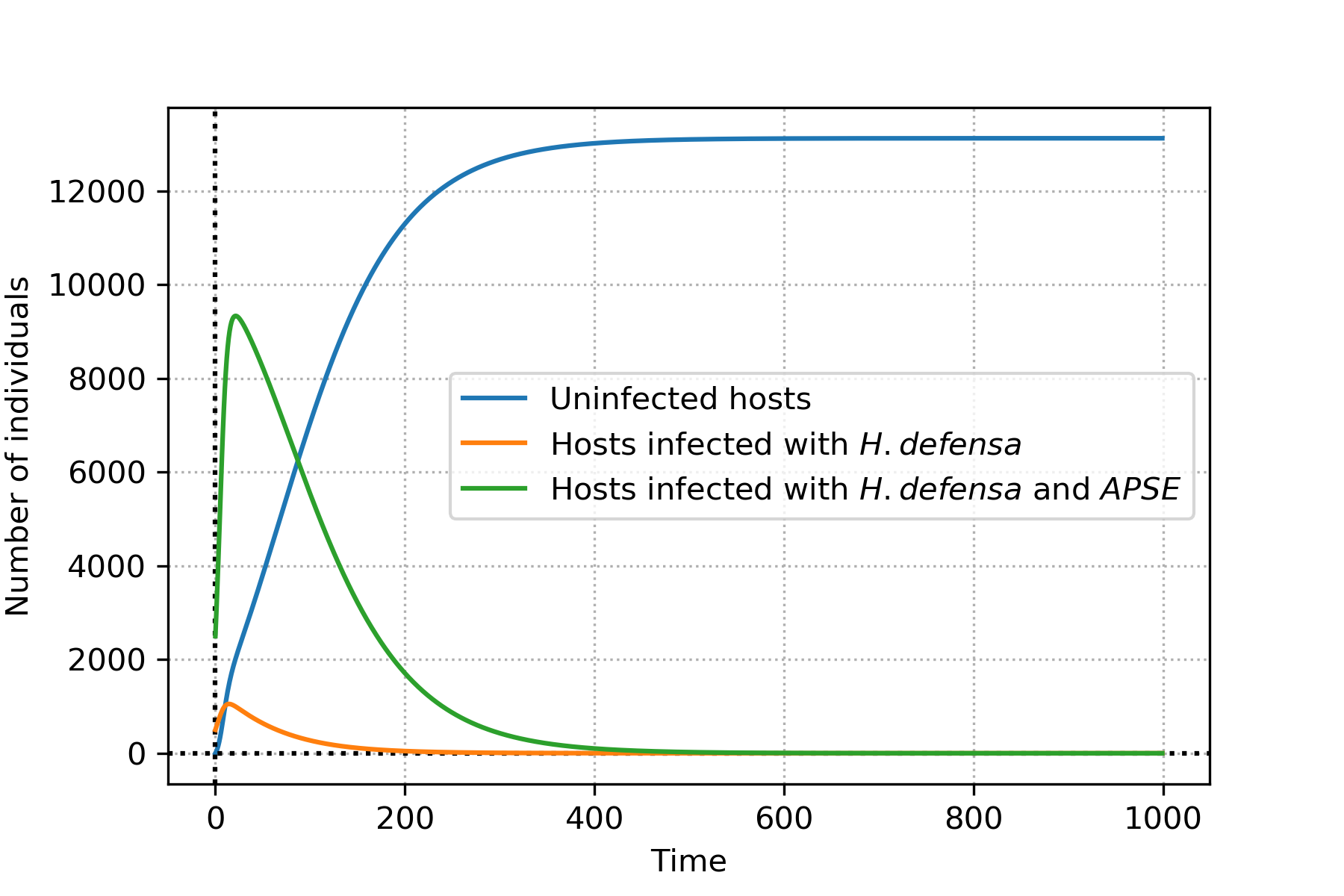}
		
		\caption{ }
	\end{subfigure}
	\begin{subfigure}{.45\textwidth}
		\centering
		\includegraphics[width=1\linewidth]{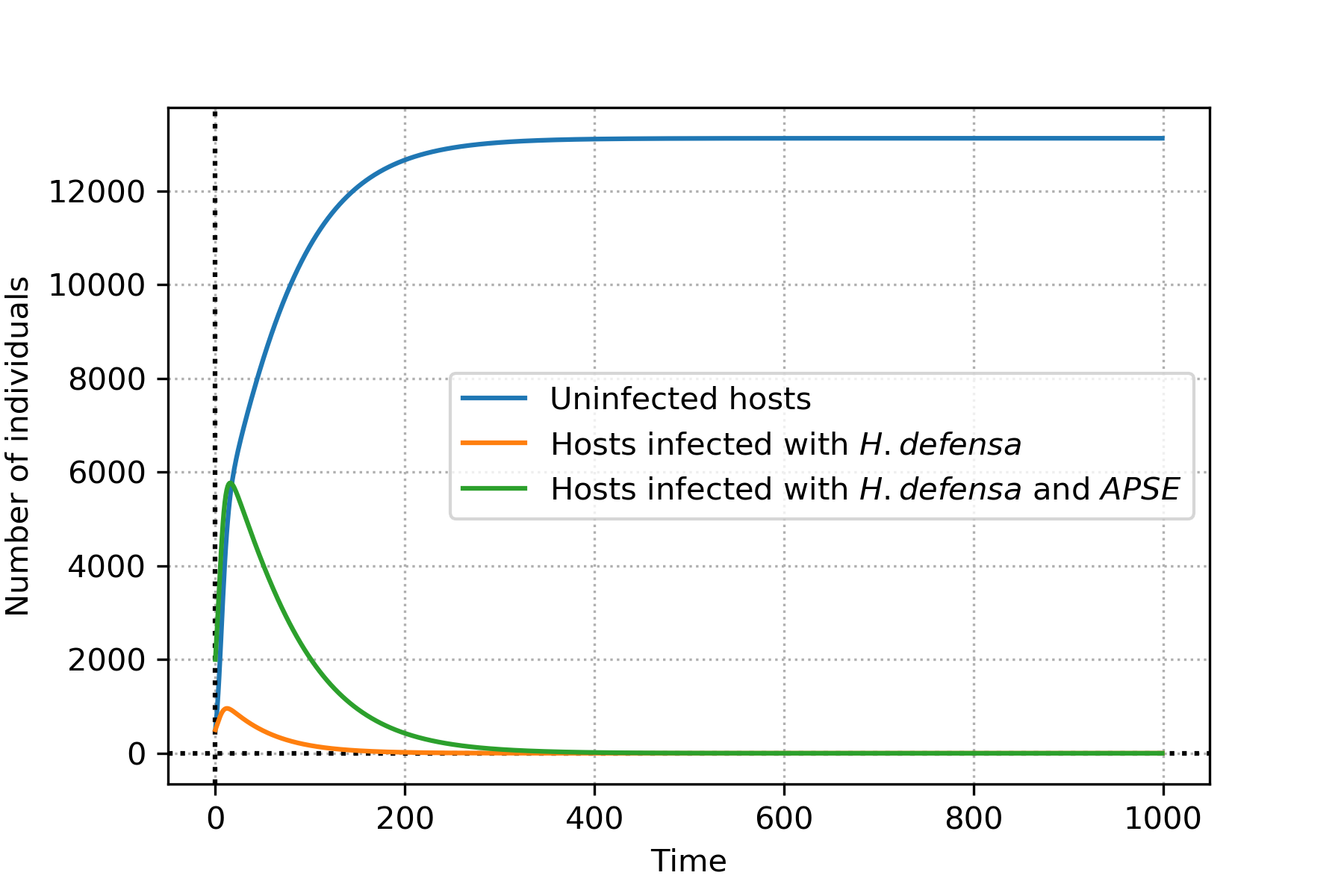}
		
		\caption{ }
		
	\end{subfigure}
	
	\caption{Population density as a function of time with parasitoid presence $P=200$ (Figures a and b) and absence $P=0$ (Figures c and d). The biotype \textit{Genista tinctoria} dynamics is represented by a and c (Parameter used as in table 1) and the biotype \textit{Medicago sativa} dynamics is represented by b and c.}

\end{figure}

\subsection{Bifurcation}

The bifurcation analysis showed that the parameter $\alpha$ have a strong influence on \textit{Genista tinctoria} and \textit{Medicago sativa} system equilibrium. As $\alpha$ increases the equilibrium point change from $\left(S=0, H=0, V=11900 \right)$ to $\left(S=12900, H=0, V=0 \right)$ in both systems. However, we observed that this inversion occurred with small values of $\alpha$ for the \textit{Genista tinctoria}. The figure 3 also shows the influence of the parameter $\theta$ that, in contrary to the previous parameter, have an expressive difference between both biotypes. As $\theta$ increases the inversion of equilibrium point from $\left(S=10900, H=0, V=0 \right)$ to $\left(S=12000, H=0, V=0 \right)$ only occurred for the \textit{M. sativa} biotype. In the other hand, for the \textit{G. tinctoria} biotype the results showed that changes at the equilibrium point only occurred for $\theta \geq 0.9$.

\begin{figure}[ht]
	\centering
	\begin{subfigure}{.45\textwidth}
		\centering
		\textit{Genista tinctoria}\par\medskip
		\includegraphics[width=1\linewidth]{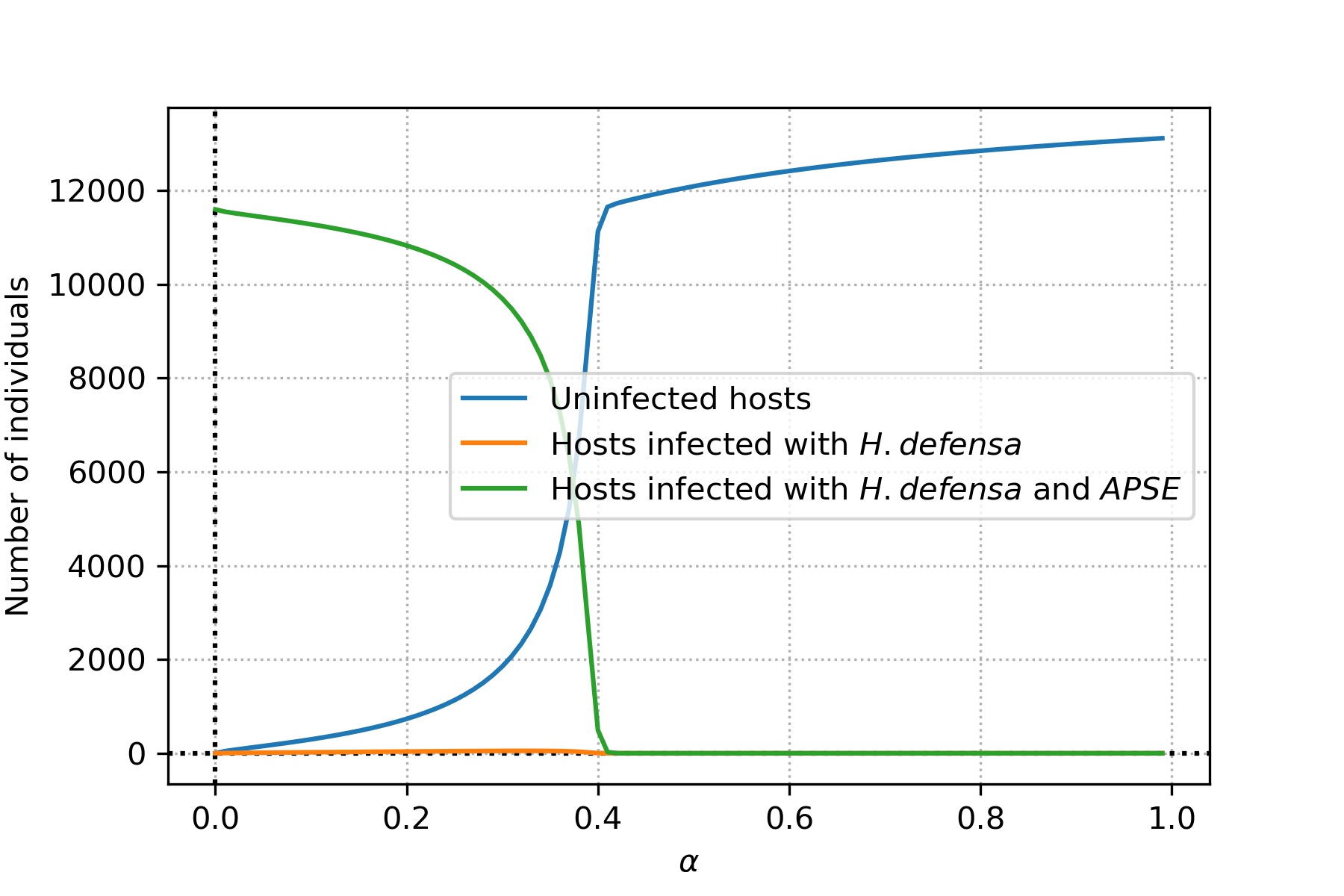}
	
		\caption{}
	\end{subfigure}
	\begin{subfigure}{.45\textwidth}
		\centering
		\textit{Medicago sativa}\par\medskip
		\includegraphics[width=1\linewidth]{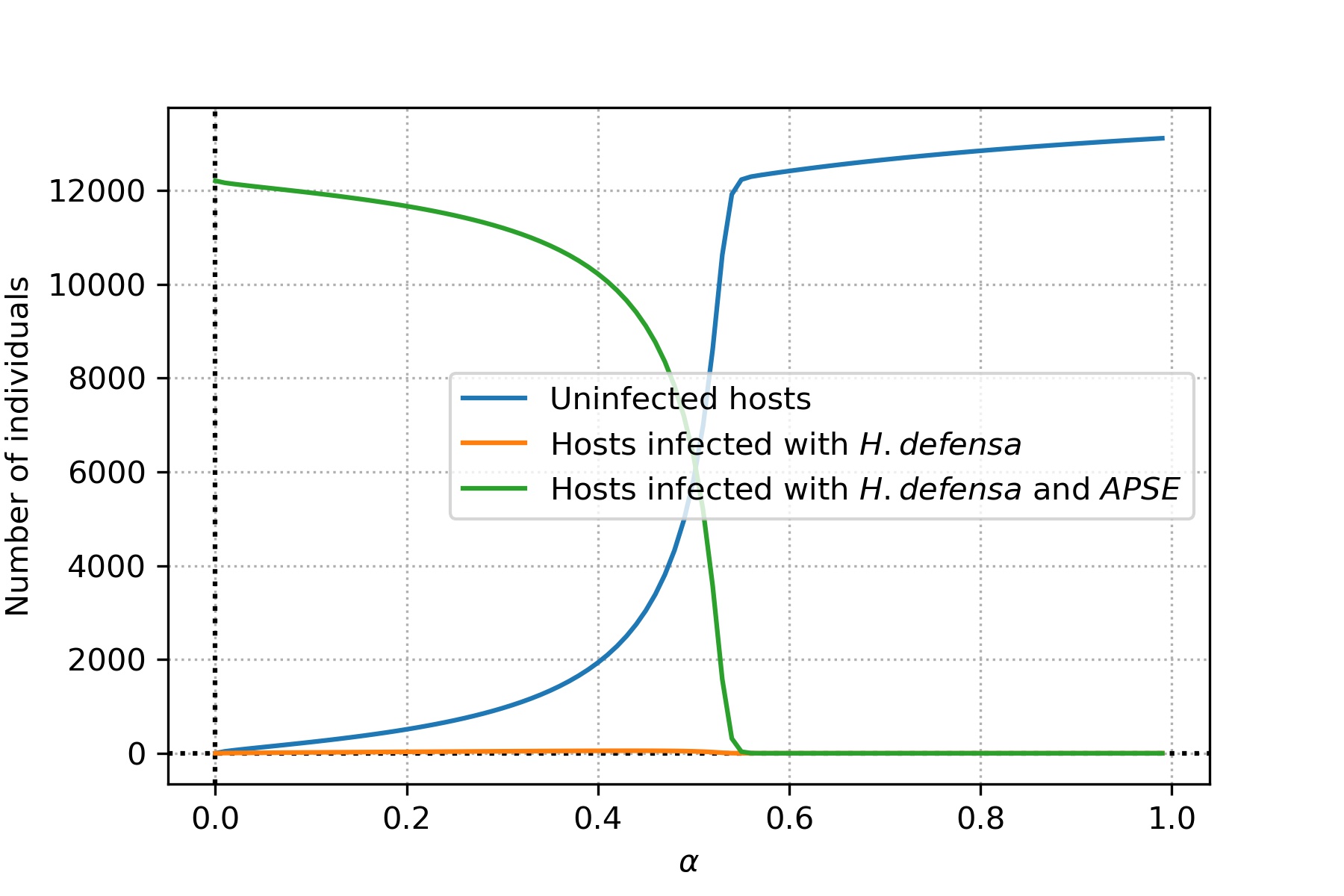}
	
		\caption{}
		
	\end{subfigure}
	
	\begin{subfigure}{.45\textwidth}
		\centering
		\includegraphics[width=1\linewidth]{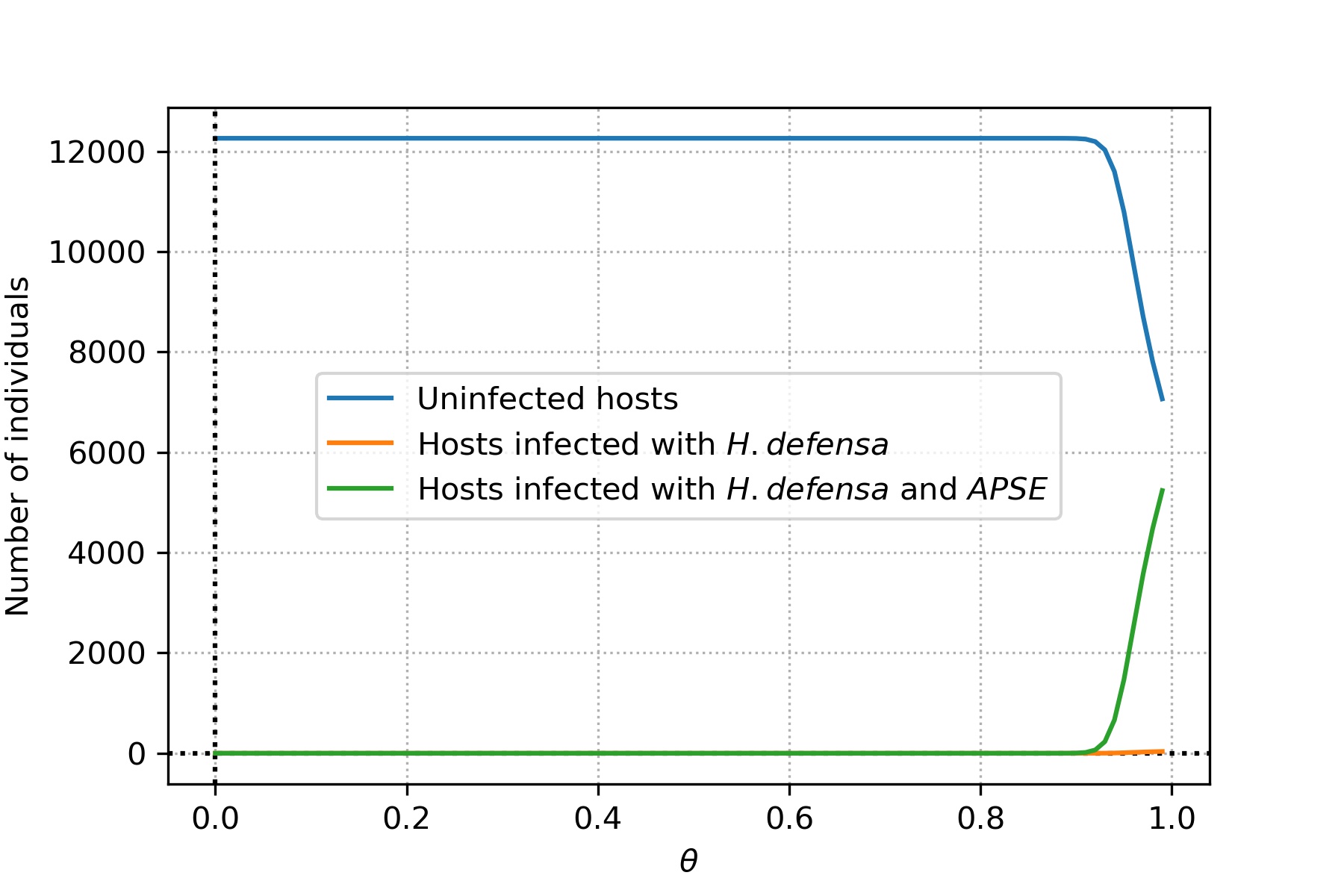}
		
		\caption{}
	\end{subfigure}
	\begin{subfigure}{.45\textwidth}
		\centering
		\includegraphics[width=1\linewidth]{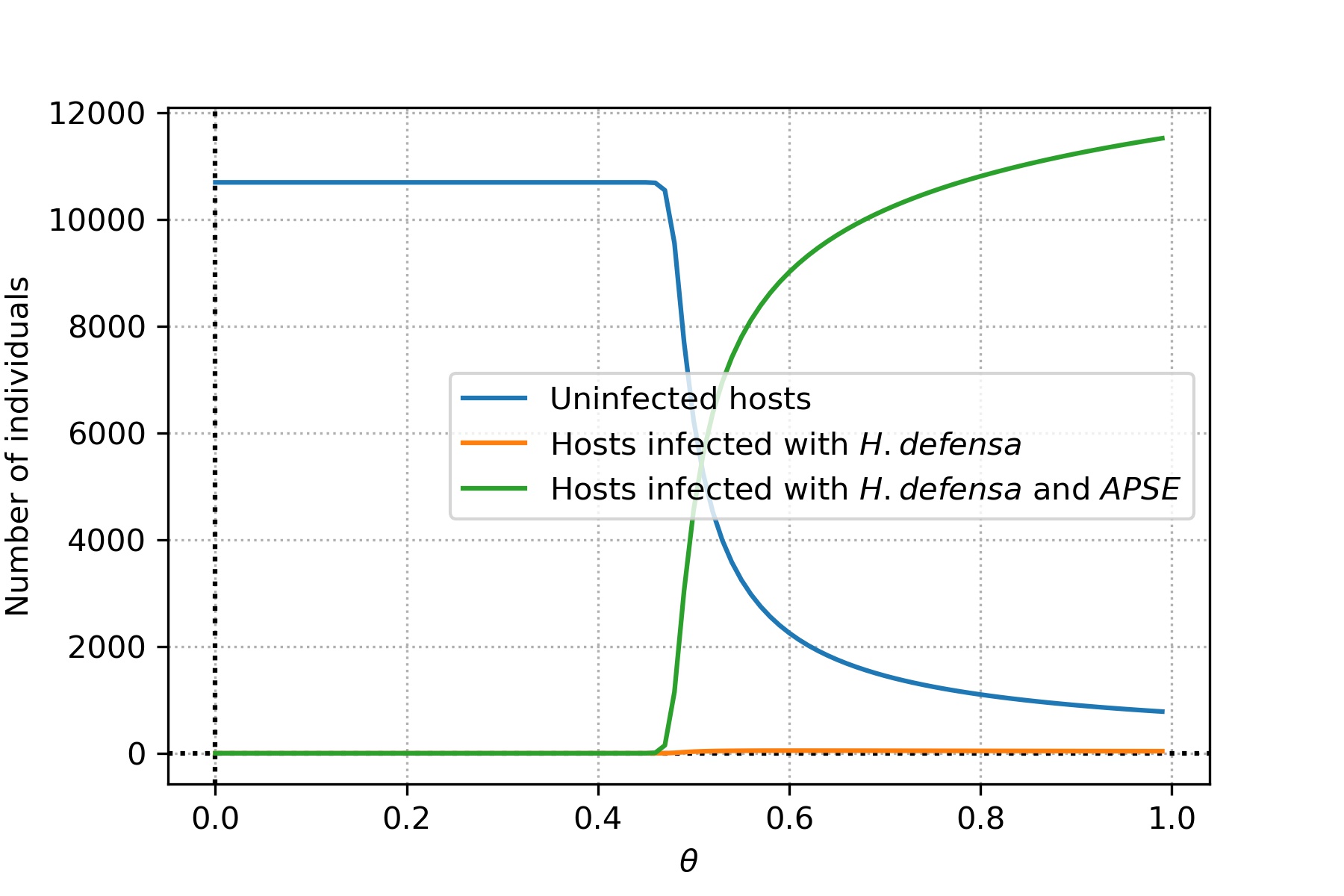}
		
		\caption{}
	\end{subfigure}

	\caption{Results of the Bifurcation analysis for the parameters $\theta$ and $\alpha$. The results are presented for both biotypes \textit{Genista tinctoria} (Letters a and c) and \textit{Medicago sativa} (Letters b and d). Here we selected the point $\left( S, H, V\right)$ after $5000$ model iterations for each parameters value of $\alpha$ and $\theta$.}

\end{figure}

Figures 4 a and c showed that as the death rate $\mu$ increases for both systems, all populations tend intuitively to zero, but the most frequent population changed between the biotypes, as previously pointed out. As the number of parasitoids $P$ increases the equilibrium point changed from $\left(S=0, H=0, V=11900 \right)$ to $\left(S=10900, H=0, V=0 \right)$ for the \textit{M. sativa} biotype. For the other system, we observed that as the parasitoid increases, this inversion does not occur for these parameter spaces.

\begin{figure}[ht]
	\centering
	\begin{subfigure}{.45\textwidth}
		\centering
		\textit{Genista tinctoria}\par\medskip
		\includegraphics[width=1\linewidth]{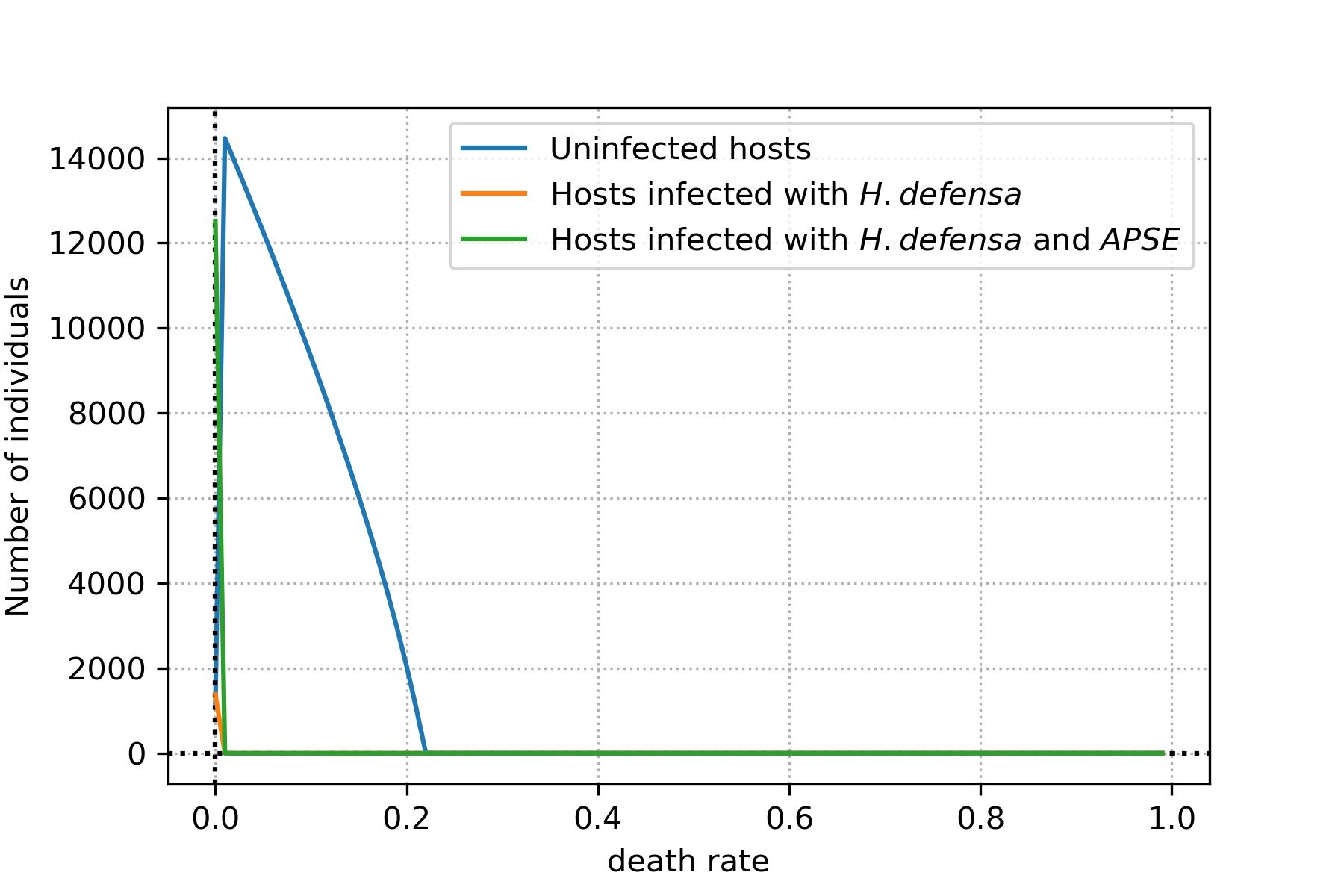}
		
		\caption{}
	\end{subfigure}
	\begin{subfigure}{.45\textwidth}
		\centering
		\textit{Medicago sativa}\par\medskip
		\includegraphics[width=1\linewidth]{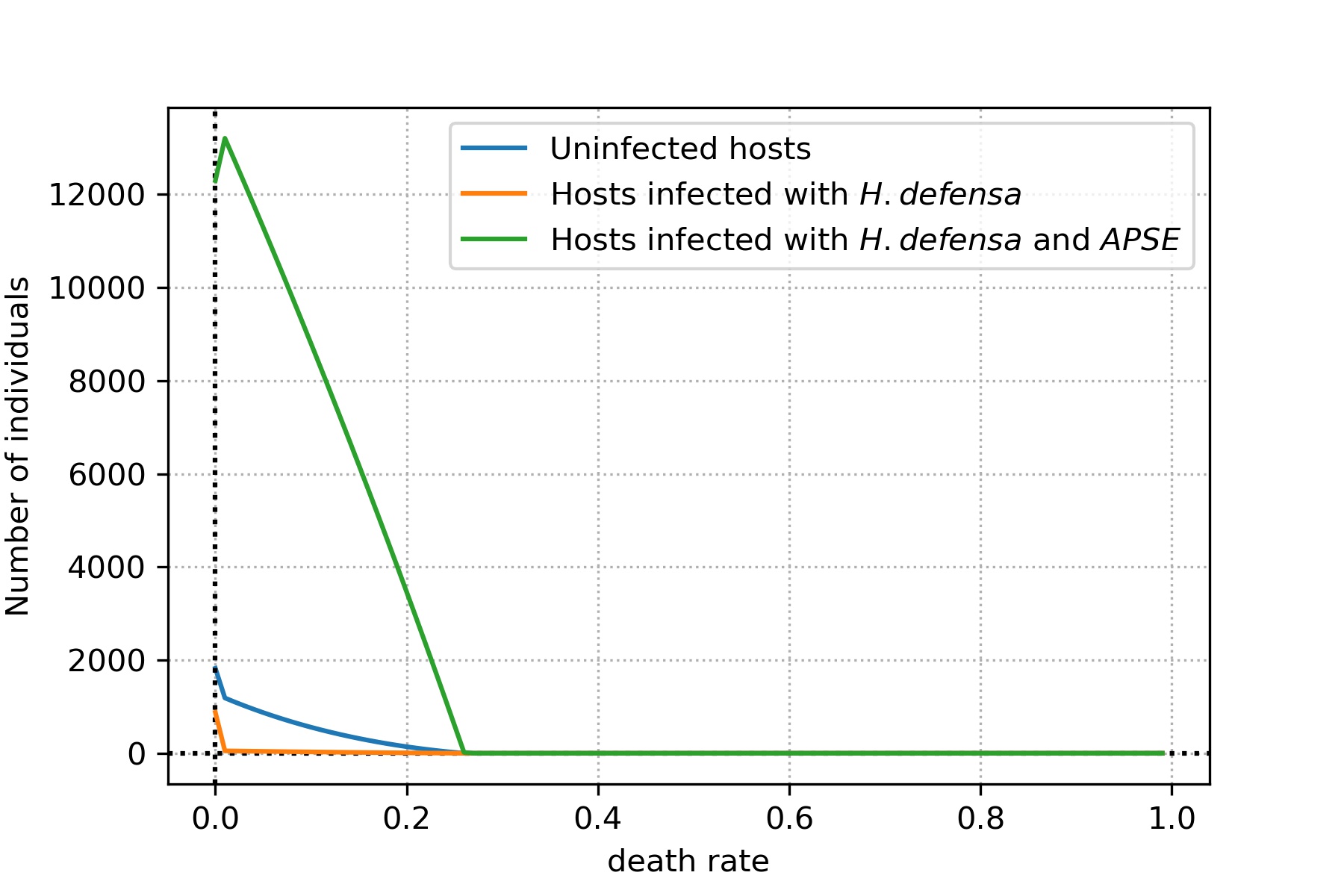}
	
		\caption{}
		
	\end{subfigure}
	
	\begin{subfigure}{.45\textwidth}
		\centering
		\includegraphics[width=1\linewidth]{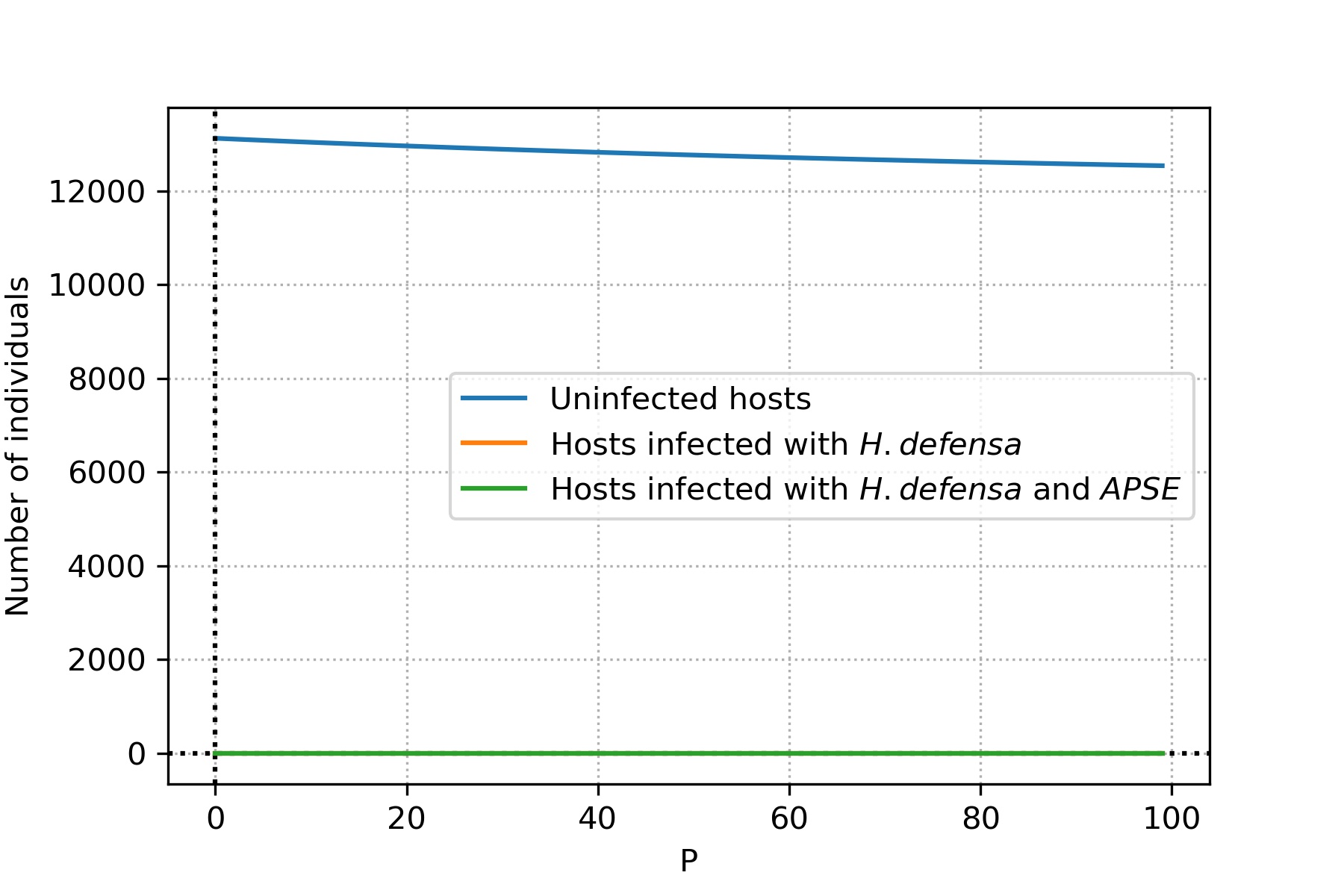}
		
		\caption{}
	\end{subfigure}
	\begin{subfigure}{.45\textwidth}
		\centering
		\includegraphics[width=1\linewidth]{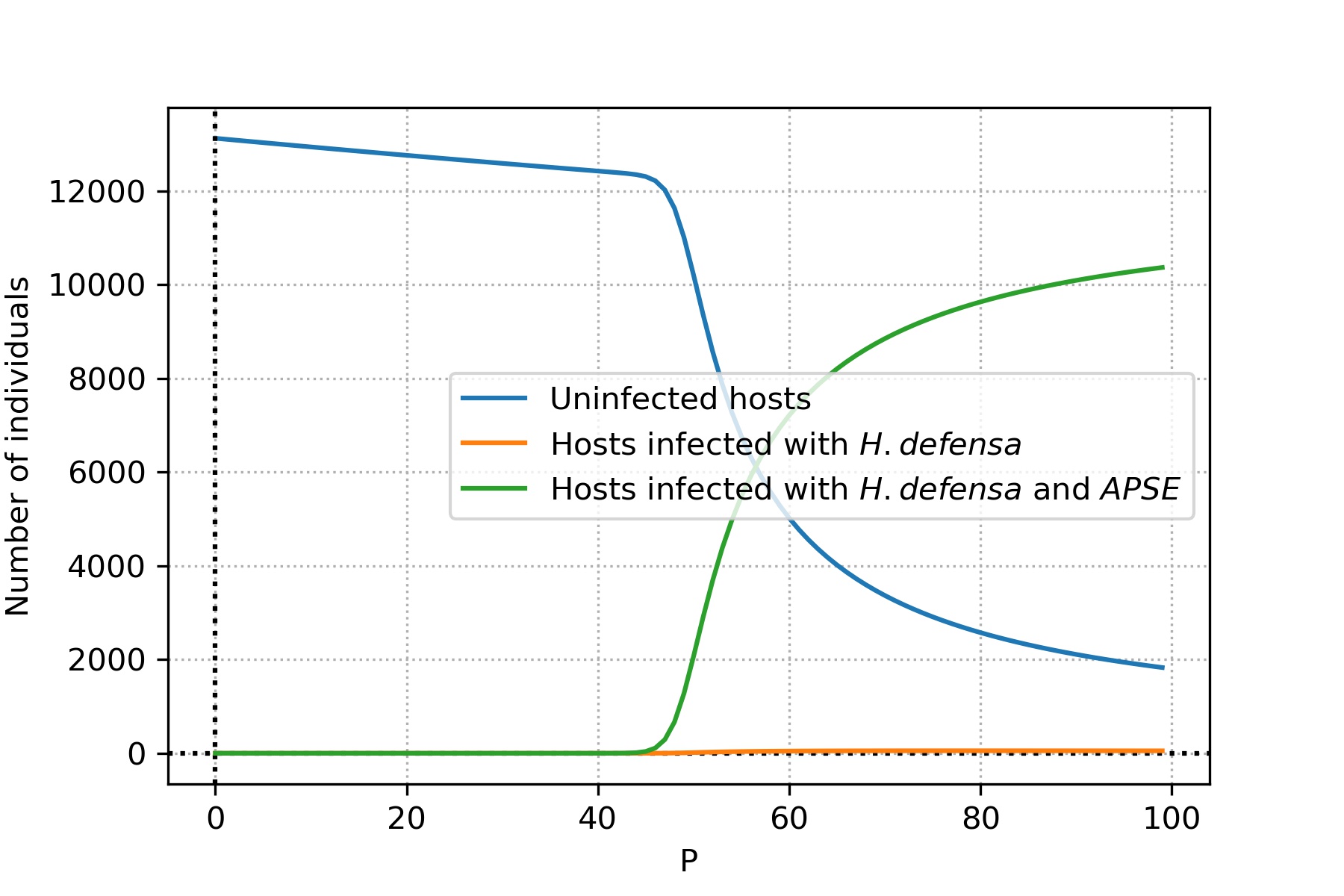}
	
		\caption{}
	\end{subfigure}

	\caption{Results of the Bifurcation analysis for the parameters $\mu$ and $P$. The results are presented for both biotypes \textit{Genista tinctoria} (Letters a and c) and \textit{Medicago sativa} (Letters b and d). Here we selected the point $\left( S, H, V\right)$ after $5000$ models iterations for each parameters value of $\mu$ and $P$.}

\end{figure}

\begin{figure}[ht]
	\centering
	\begin{subfigure}{.45\textwidth}
		\centering
		\textit{Genista tinctoria}\par\medskip
		\includegraphics[width=1\linewidth]{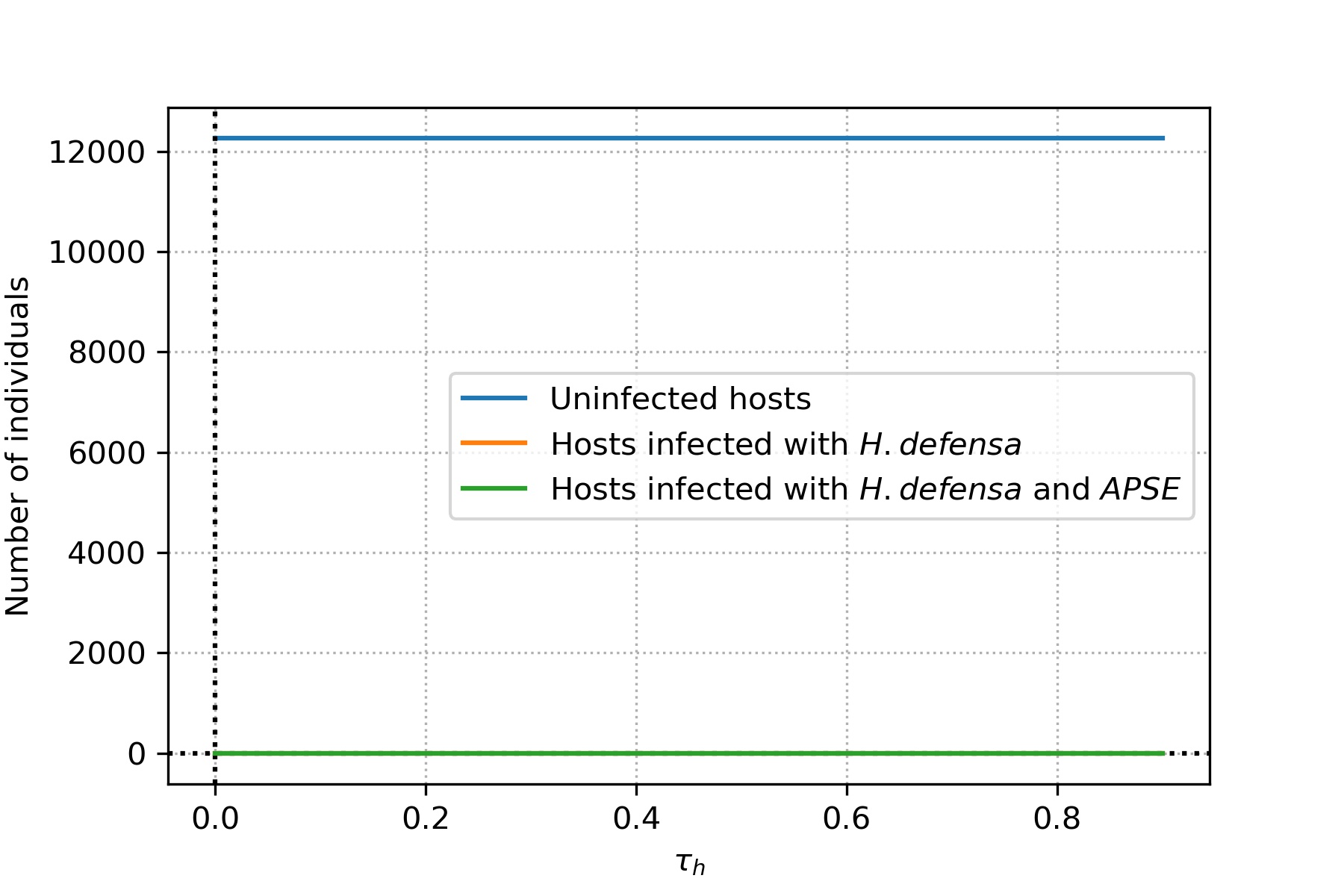}
	
		\caption{}
	\end{subfigure}
	\begin{subfigure}{.45\textwidth}
		\centering
		\textit{Medicago sativa}\par\medskip
		\includegraphics[width=1\linewidth]{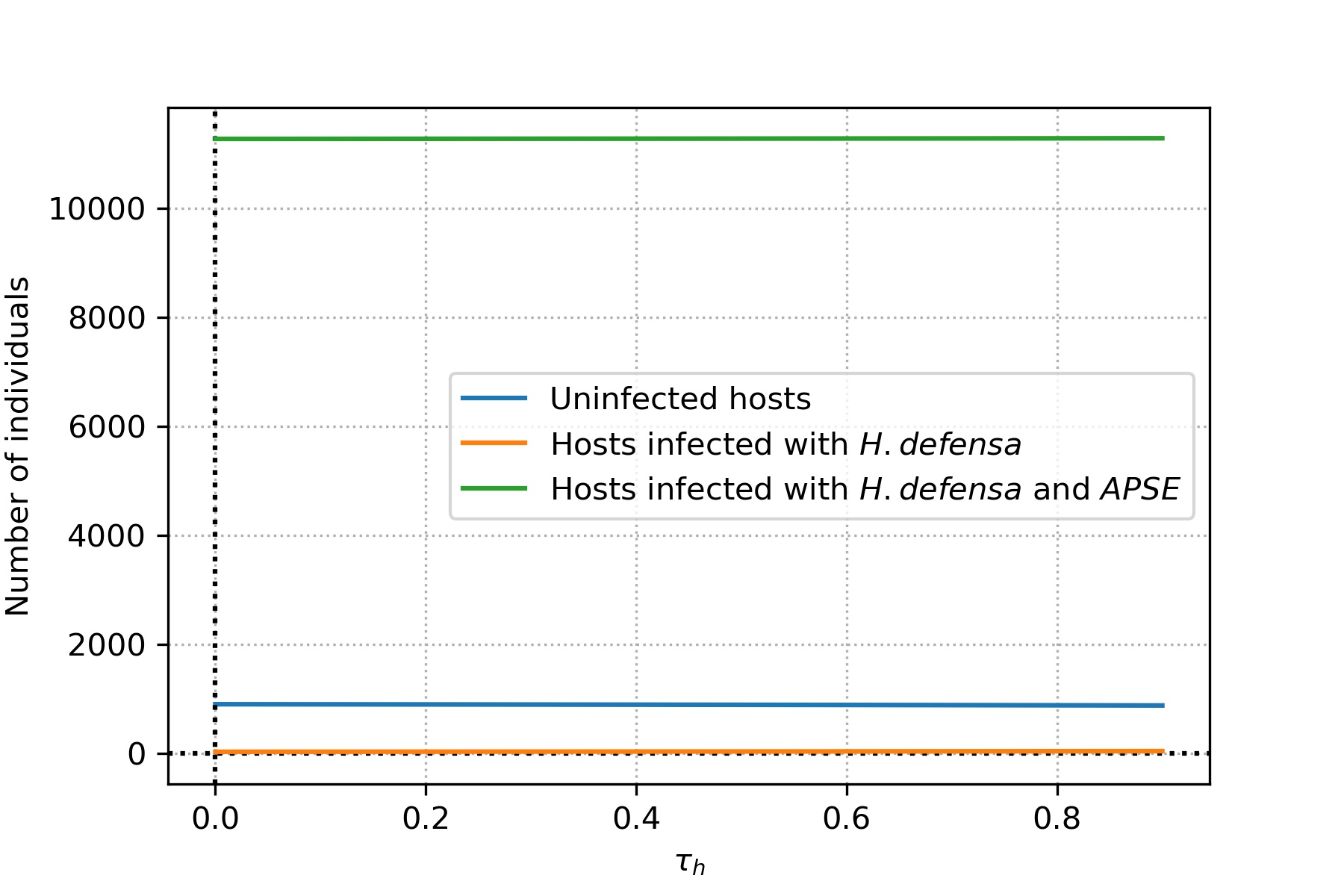}
		
		\caption{}
	\end{subfigure}
	
	\begin{subfigure}{.45\textwidth}
		\centering
		\includegraphics[width=1\linewidth]{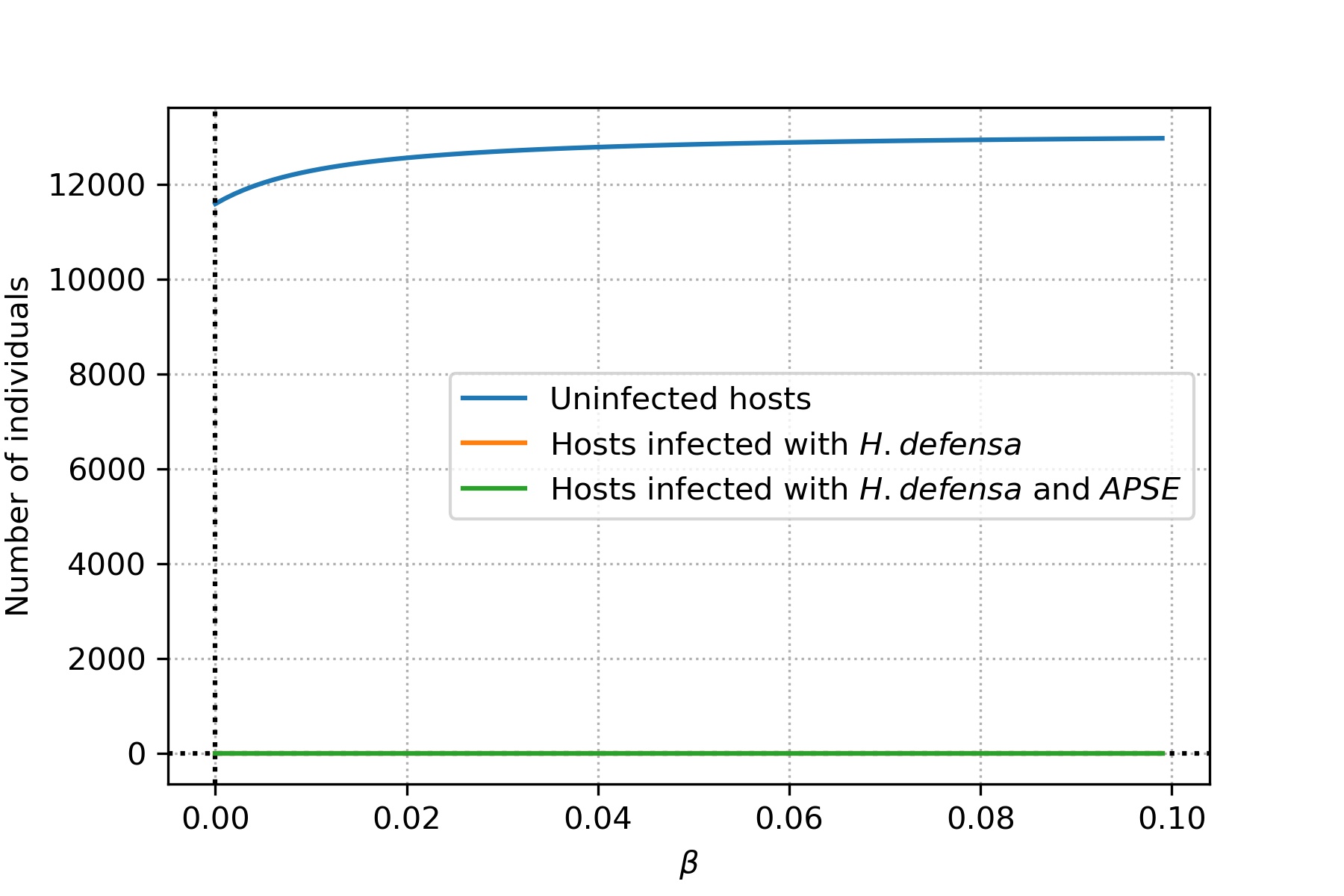}
		
		\caption{}
	\end{subfigure}
	\begin{subfigure}{.45\textwidth}
		\centering
		\includegraphics[width=1\linewidth]{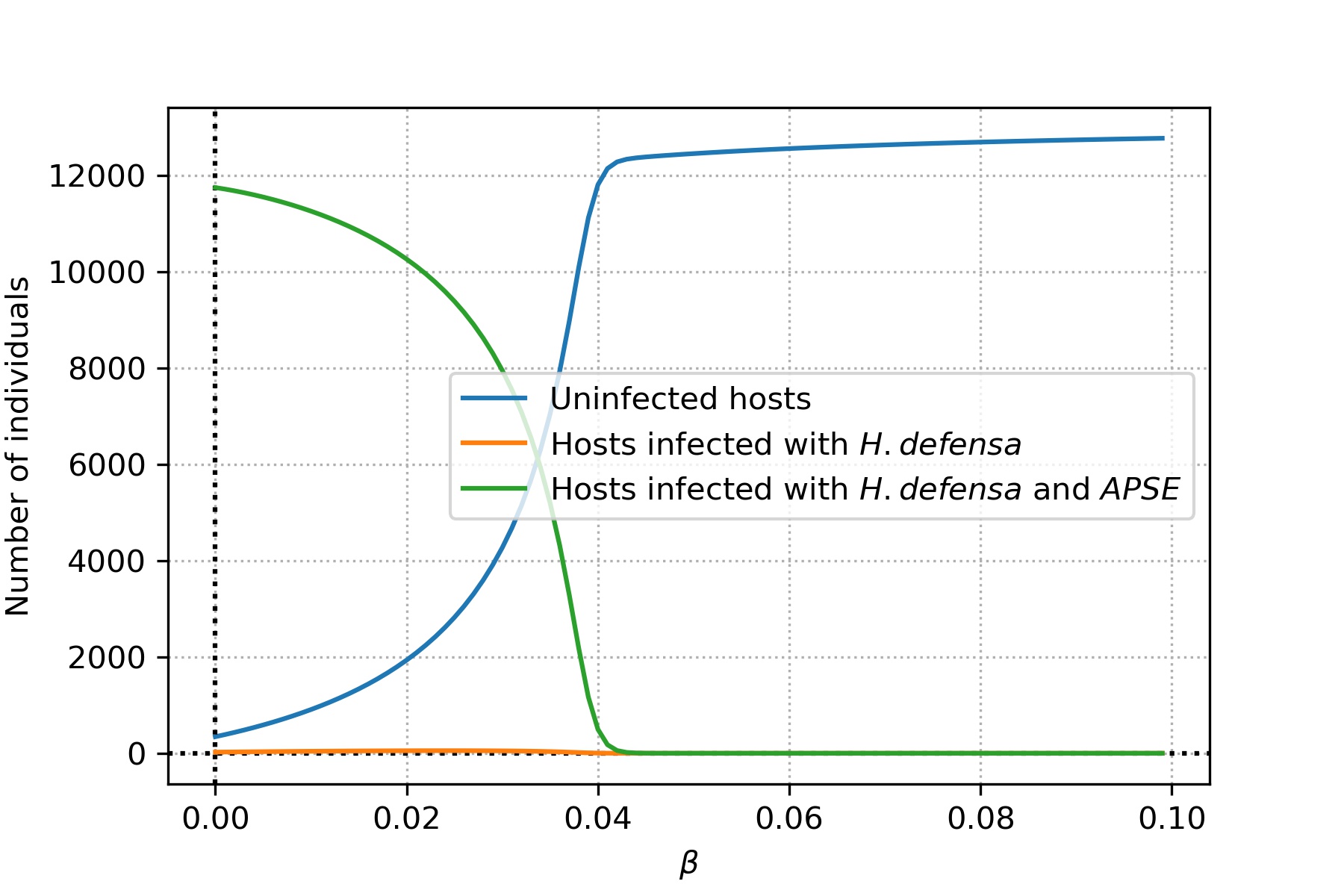}
		
		\caption{}
	\end{subfigure}

	\caption{Results of the Bifurcation analysis for the parameters $\tau_H$ and $\beta$. The results are presented for both biotypes \textit{Genista tinctoria} (Letters a and c) and \textit{Medicago sativa} (Letters b and d). Here we selected the point $\left( S, H, V\right)$ after $5000$ models iterations for each parameters value of $\tau_H$ and $\beta$.}

\end{figure}

Considering the vertical transmission of \textit{H. defensa} by the uninfected hosts $S$ ($\tau_H$) the figure 5 showed that for the biotype \textit{G. tinctoria} as this parameter increases the equilibrium point $\left( S=12200, H=0, V=0\right)$ remained the same. This behaviour also occurred for the \textit{M. sativa} biotype, but the equilibrium point was $\left( S=1000, H=0, V=11900\right)$. As $\beta$ increases, we observe the inversion of the frequent population only at the \textit{M. sativa} biotype. This inversion occurred from the equilibrium point $\left( S=100, H=0, V=11900\right)$ to $\left( S=12200, H=0, V=0\right)$. On the contrary, this inversion does not occur at the biotype \textit{G. tinctoria}, given that as $\beta$ increases, the equilibrium with the uninfected hosts $S$ as the most frequent population remains the same.

\begin{figure}[ht]
	\centering
	\begin{subfigure}{.45\textwidth}
		\centering
		\textit{Genista tinctoria}\par\medskip
		\includegraphics[width=1\linewidth]{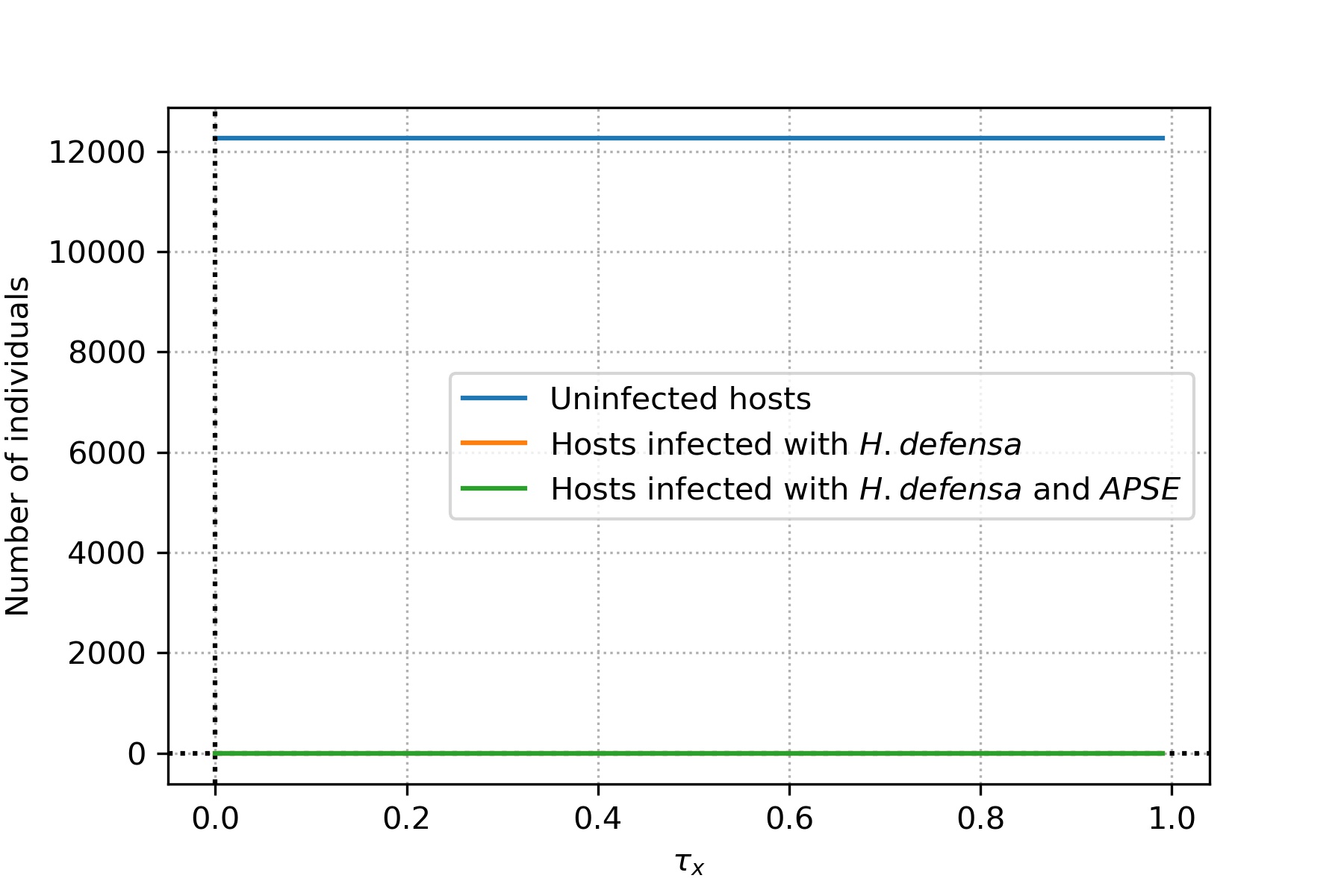}
		
	\end{subfigure}
	\begin{subfigure}{.45\textwidth}
		\centering
		\textit{Medicago sativa}\par\medskip
		\includegraphics[width=1\linewidth]{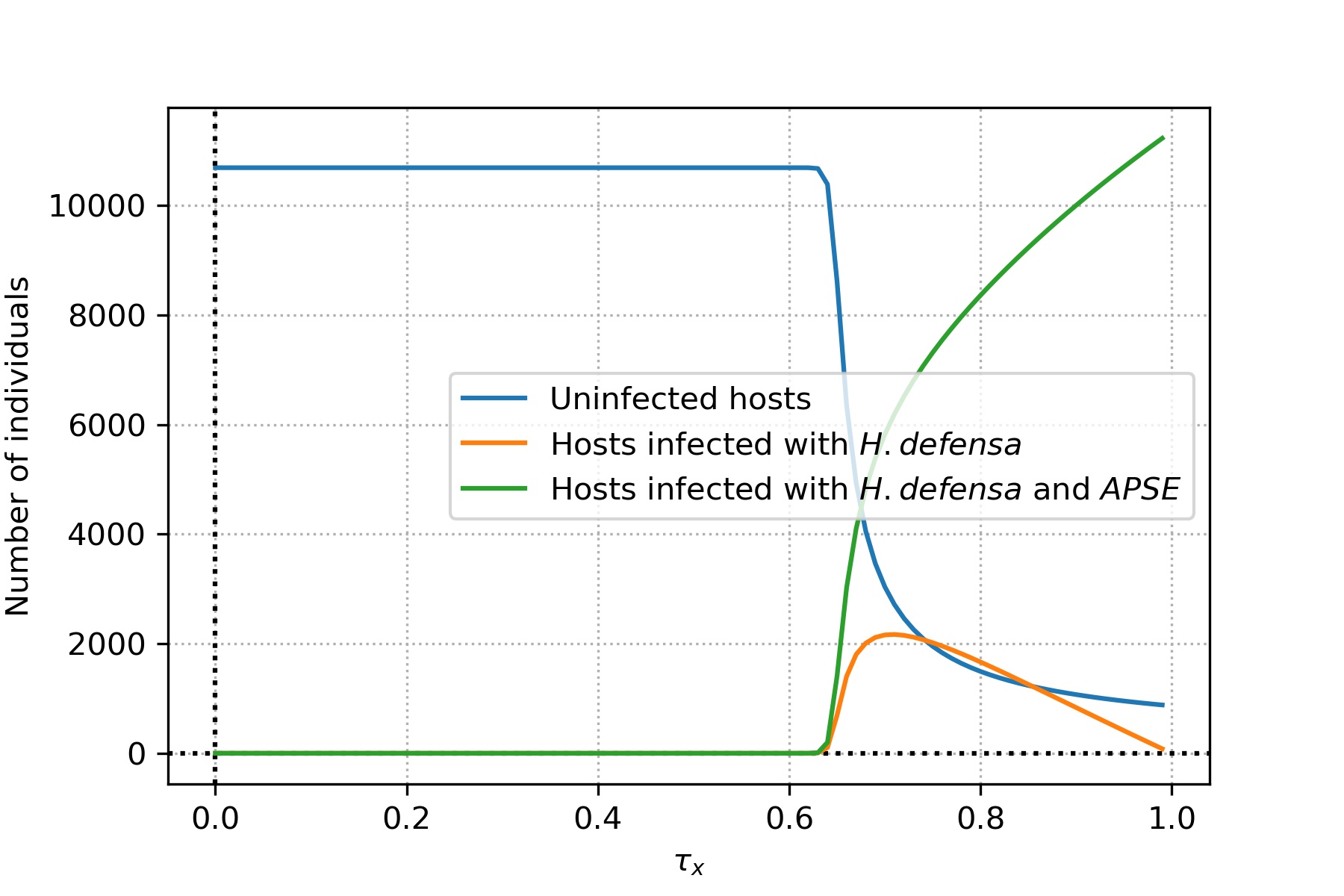}

	\end{subfigure}
	
	\begin{subfigure}{.45\textwidth}
		\centering
		\includegraphics[width=1\linewidth]{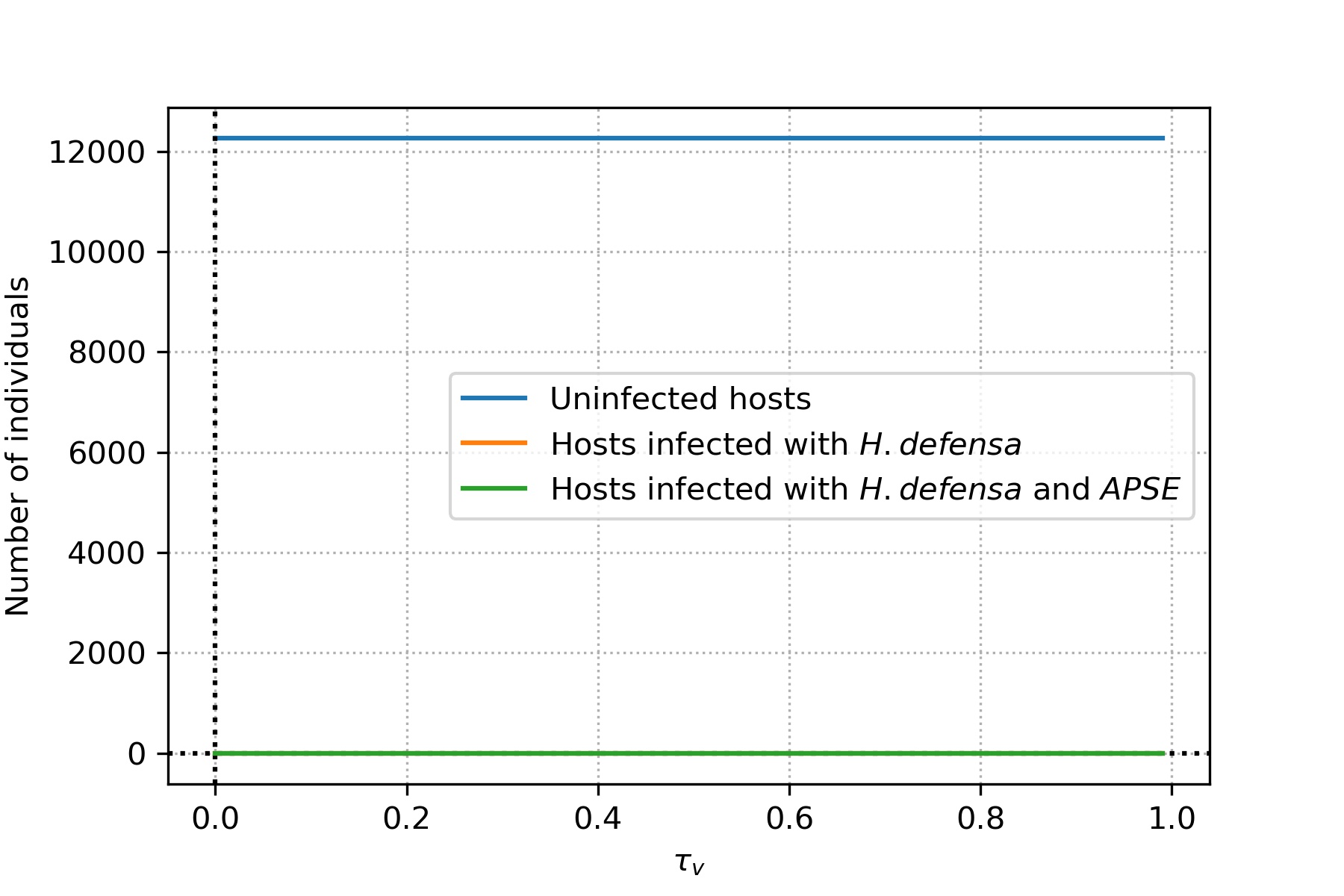}
		
	\end{subfigure}
	\begin{subfigure}{.45\textwidth}
		\centering
		\includegraphics[width=1\linewidth]{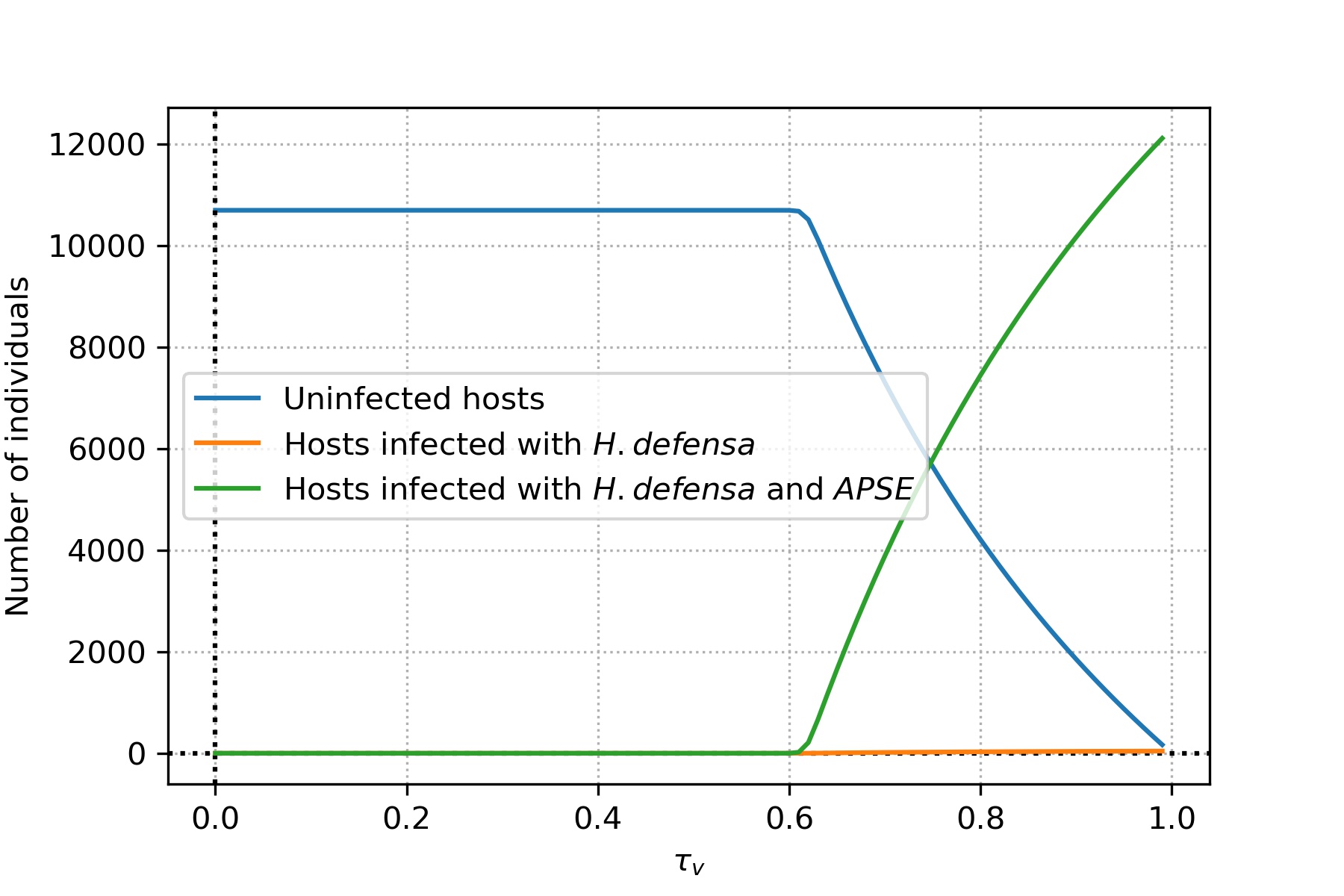}

	\end{subfigure}

	\caption{Results of the Bifurcation analysis for the parameters $\tau_X$ and $\tau_V$. The results are presented for both biotypes \textit{Genista tinctoria} (Letters a and c) and \textit{Medicago sativa} (Letters b and d). Here we selected the point $\left( S, H, V\right)$ after $5000$ models iterations for each parameters value of $\tau_X$ and $\tau_V$.}

\end{figure}

Considering the vertical transmission of \textit{H. defensa} by hosts infected with \textit{H. defensa} plus \textit{APSE} ($\tau_V$), for the biotype \textit{M. sativa} (Figure 6) as this parameter increases the equilibrium point change from $\left( S=10900, H=0, V=0\right)$ to $\left( S=0, H=0, V=12000\right)$. However, for the biotype \textit{G. tinctoria} as $\tau_V$ increases the equilibrium point $\left( S=12100, H=0, V=0\right)$ remains the same. The vertical transmission of \textit{APSE} had similar behaviour for both scenarios, but for the biotype \textit{M. sativa} as $\tau_X$ increases the system presented an equilibrium with all population, for example $\left( S=2000, H=2000, V=7000\right)$, showing the possibility of coexistence of these hosts. Finally, for the biotype \textit{G. tinctoria} as $\tau_X$ increases the equilibrium point $\left( S=1200, H=0, V=0\right)$ remains constant. 

\begin{figure}[ht]
	\centering
	
	\begin{subfigure}{.45\textwidth}
		\centering
		\textit{Genista tinctoria}\par\medskip
		\includegraphics[width=1\linewidth]{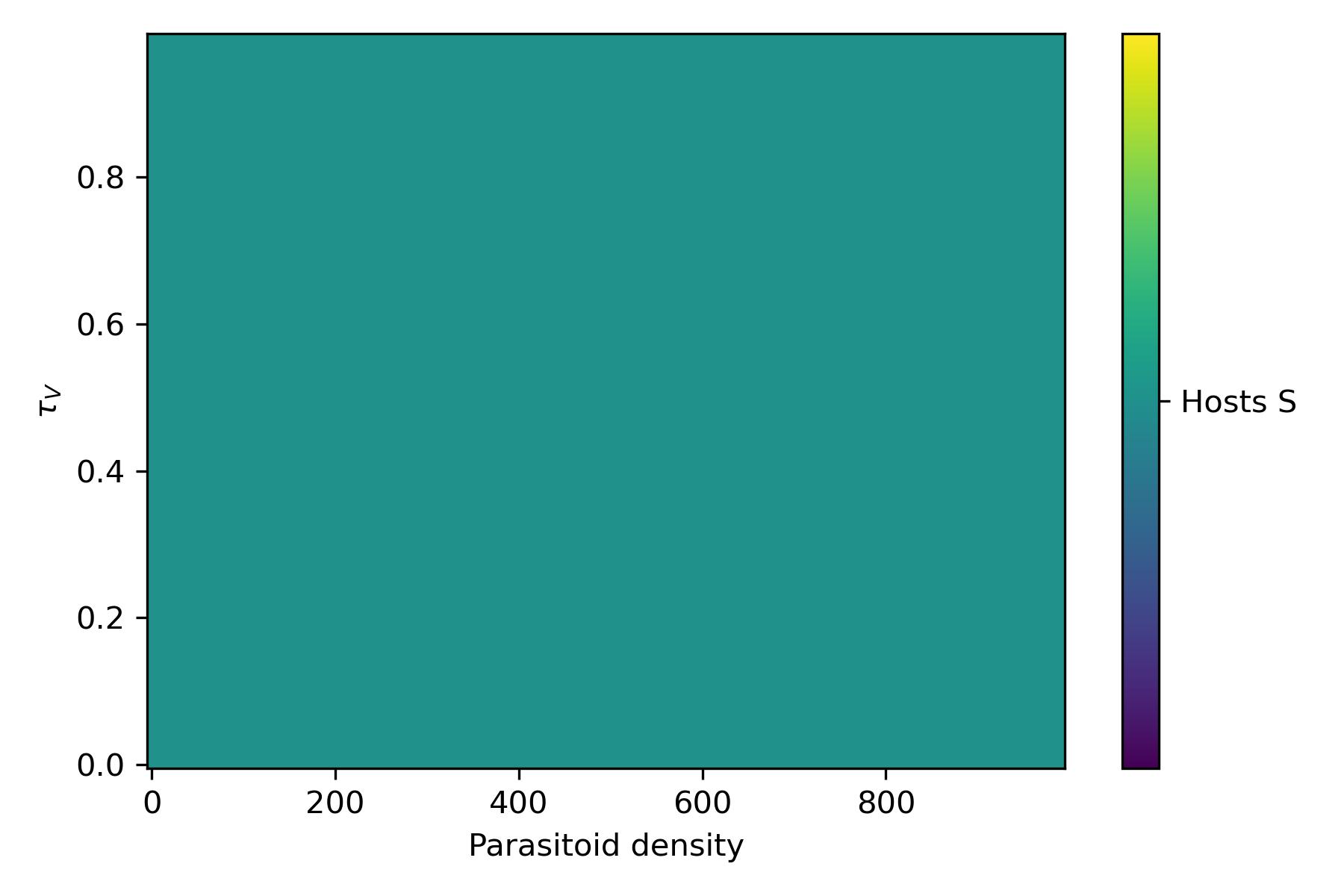}
		
		\caption{}
	\end{subfigure}
	\begin{subfigure}{.45\textwidth}
		\centering
		\textit{Medicago sativa}\par\medskip
		\includegraphics[width=1\linewidth]{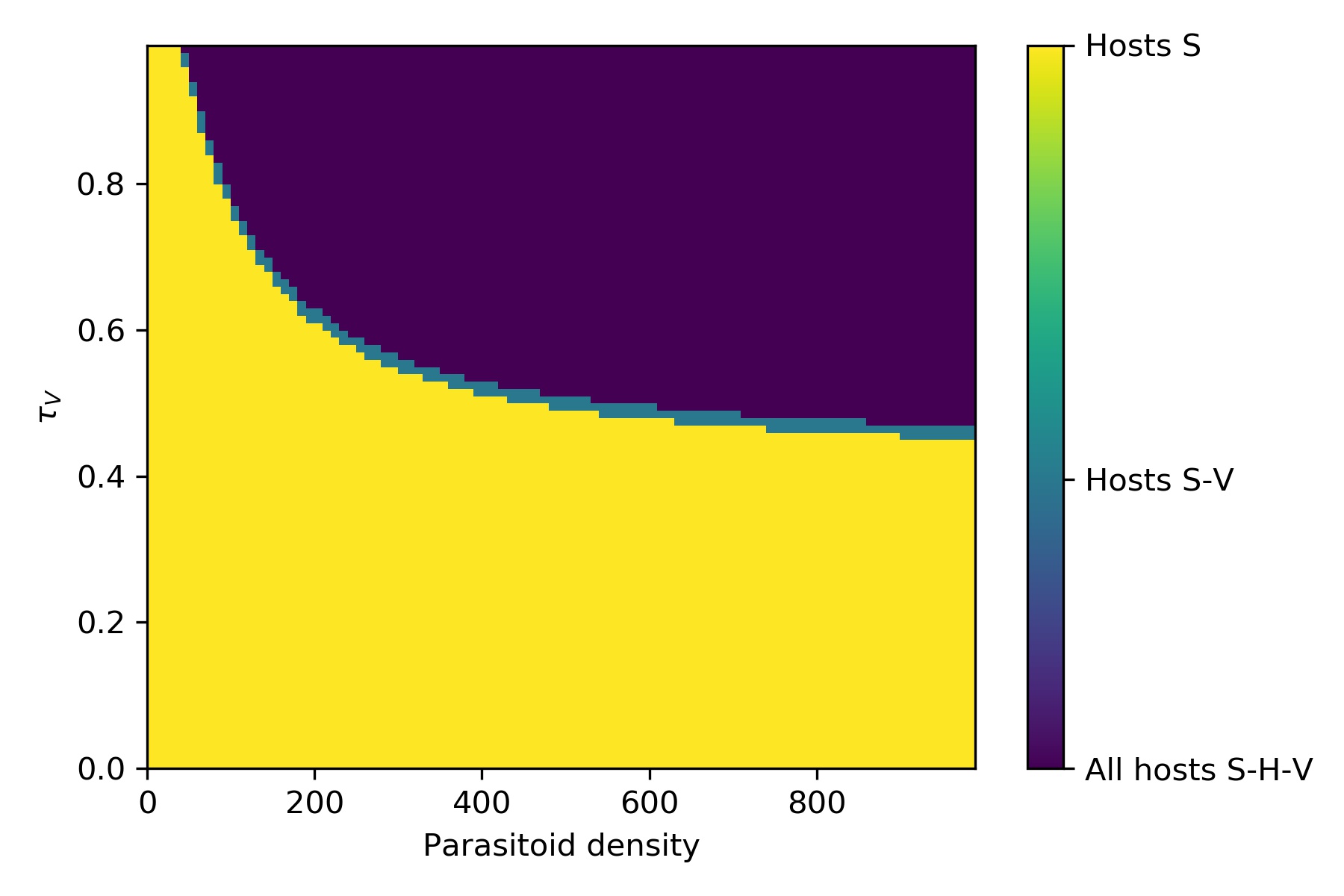}
		
		\caption{}
	\end{subfigure}
	\begin{subfigure}{.45\textwidth}
		\centering
		\includegraphics[width=1\linewidth]{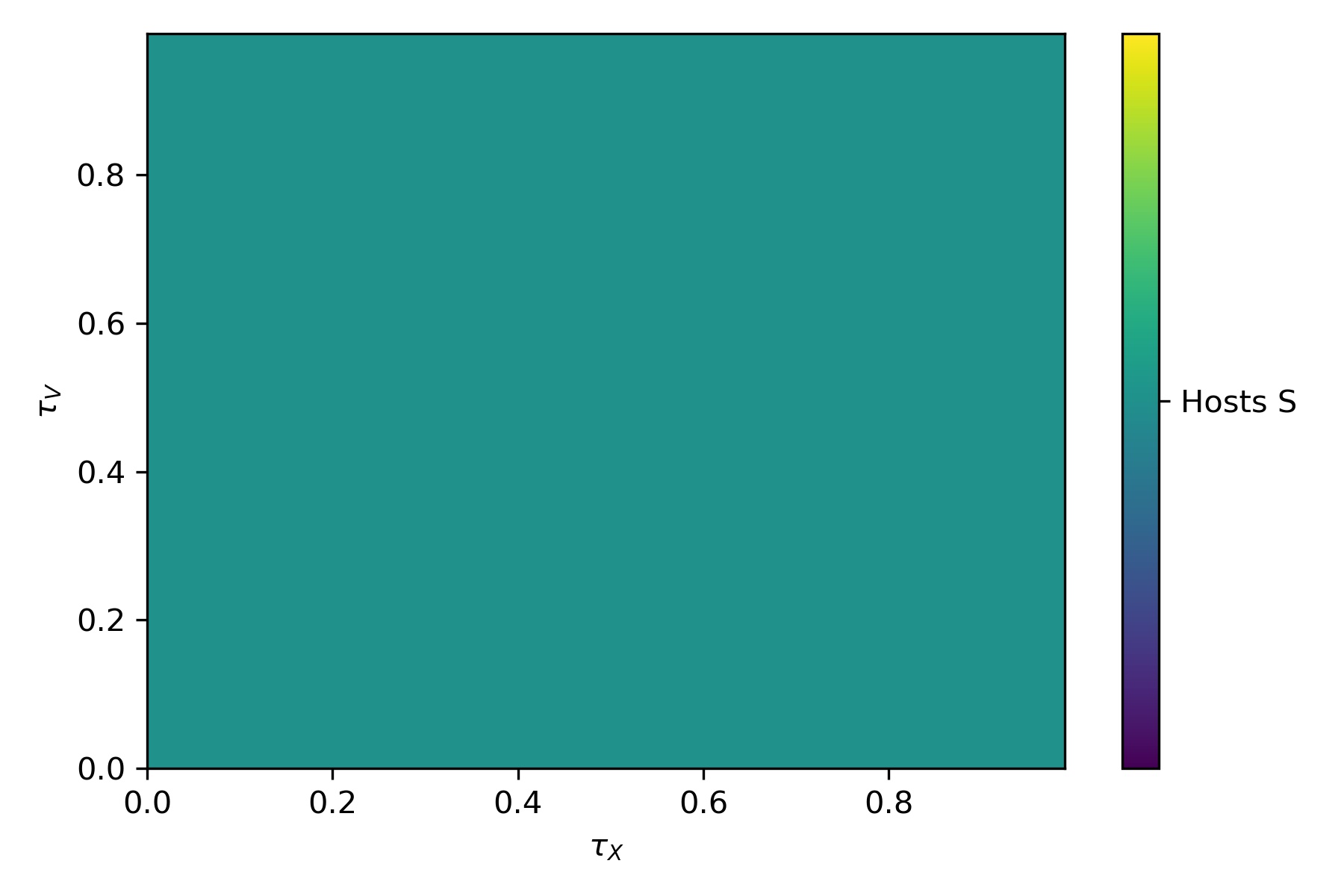}
		
		\caption{}
	\end{subfigure}
	\begin{subfigure}{.45\textwidth}
		\centering
		\includegraphics[width=1\linewidth]{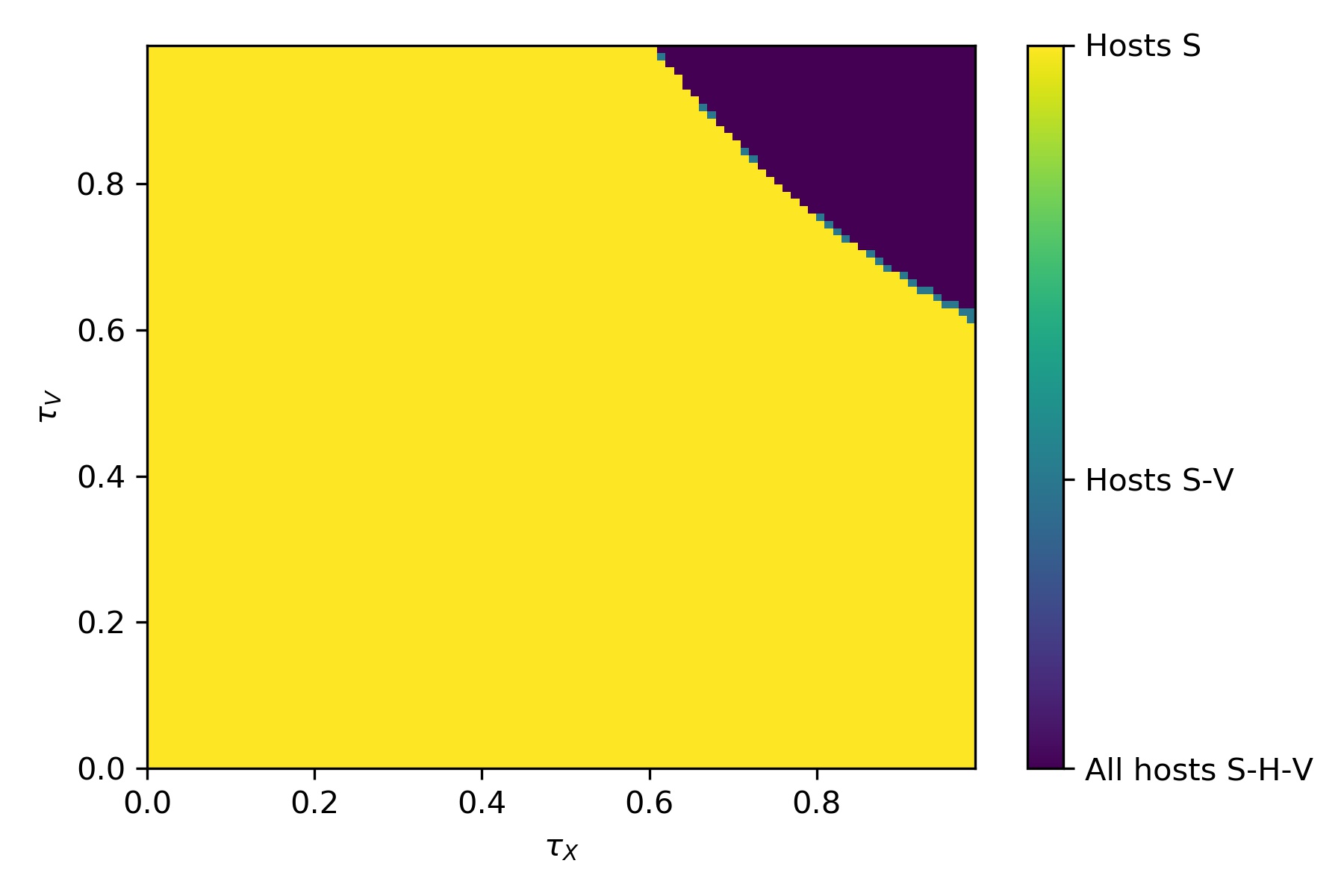}
		
		\caption{}
		
	\end{subfigure}
	
	\begin{subfigure}{.45\textwidth}
		\centering
		\includegraphics[width=1\linewidth]{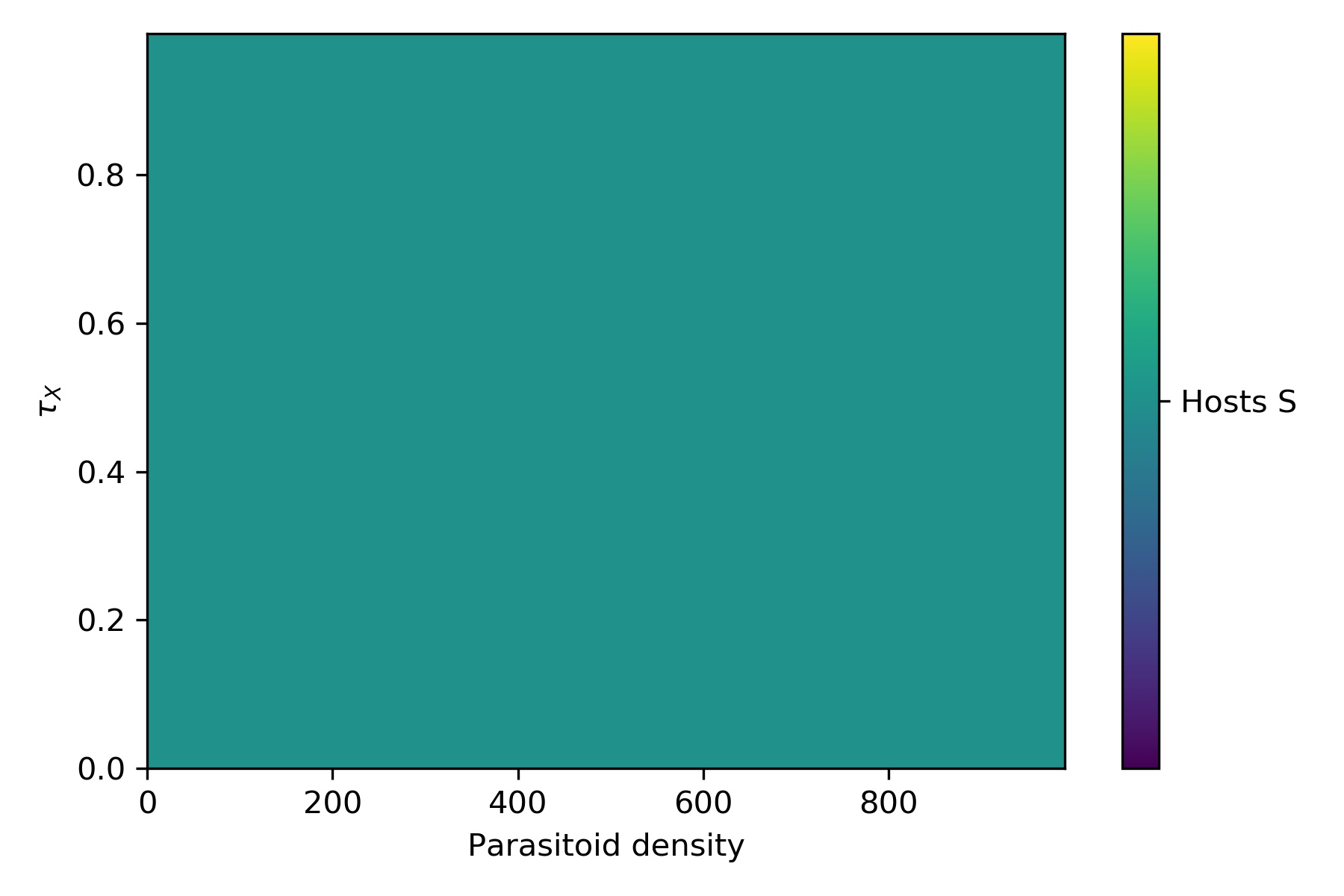}
		
		\caption{}
	\end{subfigure}
	\begin{subfigure}{.45\textwidth}
		\centering
		\includegraphics[width=1\linewidth]{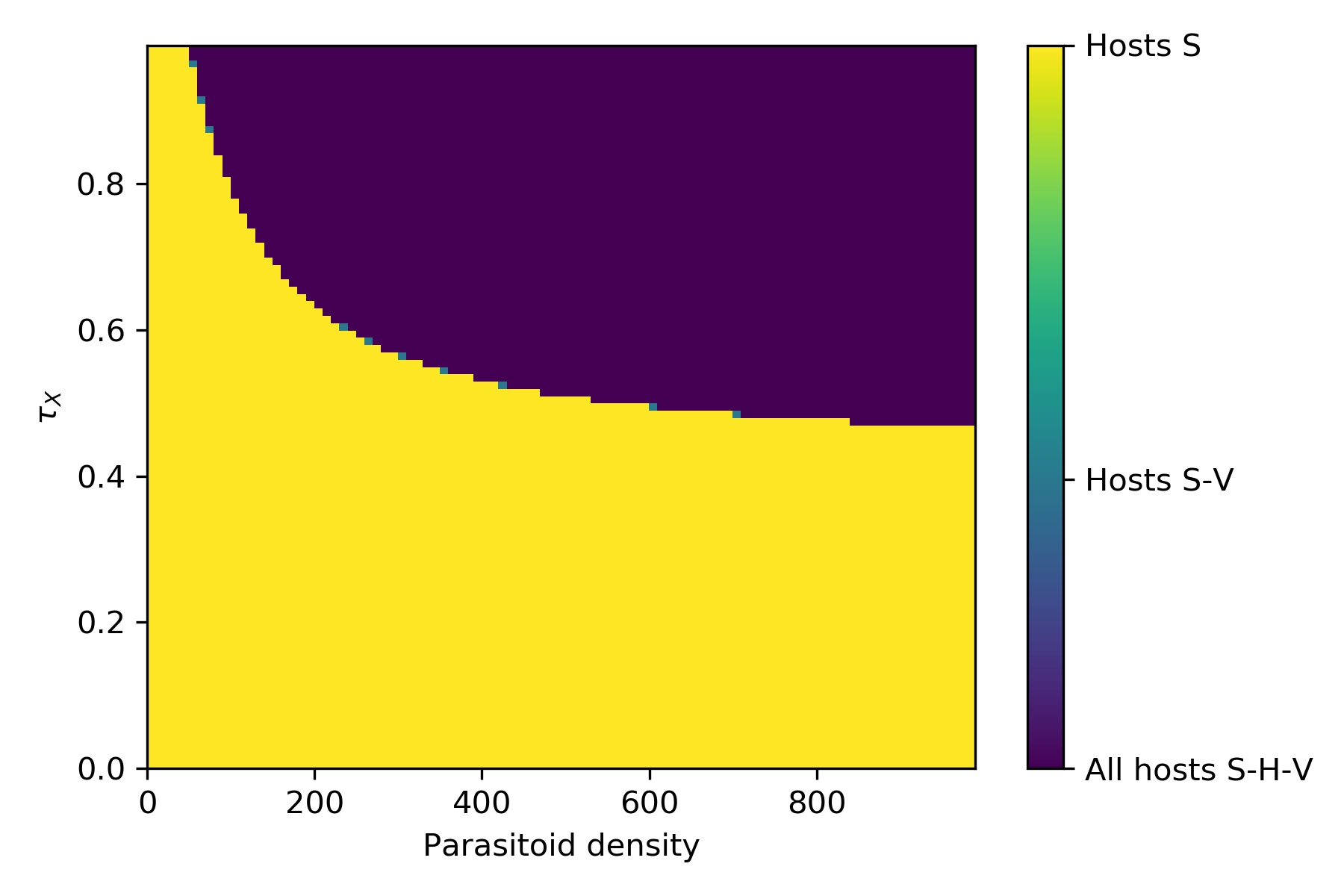}
		
		\caption{}
	\end{subfigure}

	\caption{Population persistence for a range of parasitism ($P$), vertical transmission of the bacteriophage ($\tau_X$) and vertical transmission of \textit{H. defensa} by $V$ considering the \textit{Genista tinctoria} (a, c and e) and \textit{Medicago sativa} biotype (b, d and f).}

\end{figure}

Based on the results of the population persistence in figure 7, the grids performed with vertical transmission of \textit{APSE} ($\tau_X$), vertical transmission of \textit{H. defensa} by $V$ ($\tau_V$) and parasitoid density $P$ did not change the \textit{Genista tinctoria} system equilibrium, given that the uninfected hosts $S$ persisted as the unique population regardless of the parameter's values. On the contrary, we observed that the \textit{Medicago sativa} system varies its composition according to the grid of these parameters. The second column of figure 7 shows that the grids resulted in the following population combinations:
\begin{enumerate}
    \item Uninfected hosts ($S$)
    \item Uninfected hosts ($S$) and infected hosts with \textit{H. defensa} plus \textit{APSE} ($V$)
    \item Uninfected hosts ($S$), infected hosts with \textit{H. defensa} and infected hosts with \textit{H. defensa} plus \textit{APSE} ($V$),
\end{enumerate}
Figure 7 b and f showed that according to the grid ($\tau_X$, $P$) and ($\tau_V$, $P$) the uninfected hosts ($S$), infected hosts with \textit{H. defensa} ($H$) and infected hosts with \textit{H. defensa} plus \textit{APSE} ($V$) are more frequent than the grid of $\tau_X$ and $\tau_V$ presented in figure d. 



\subsection{Protection evolution}

Figure 10 shows that after $5000$ model iterations with the presence of $P=200$ parasitoids, the \textit{G. tinctoria} system persisted with only uninfected hosts. Thus, according to the population density, bifurcation analysis and this figure, we observed that the selected parameter represents the \textit{G. tinctoria} system stabilize with $\left( S > 0, H=0, V=0\right)$ regardless of the presence or absence of parasitoids. Given this result, we only considered the biotype \textit{M. sativa} for this analysis. Finally, the figure 11 shows that the transmission of \textit{APSE} ($\tau_X$) and the transmission of \textit{H. defensa} by $V$ ($\tau_V$) are proportional to the time that hosts infected with \textit{H. defensa} plus \textit{APSE} reaches $0$. This time vary from $0$ to $700$ model iterations for the parameter $\tau_V$ and a variation from $0$ to $500$ for the parameter $\tau_X$, where $0$ indicates that the presence of parasitoids did not influence the increase of the infected hosts $V$. We observed that the time only increased when $\tau_V\geq 0.6$ and $\tau_X> 0.6$.
\begin{figure}
    \centering
    \begin{subfigure}{.45\textwidth}
		\centering
		\includegraphics[width=1\linewidth]{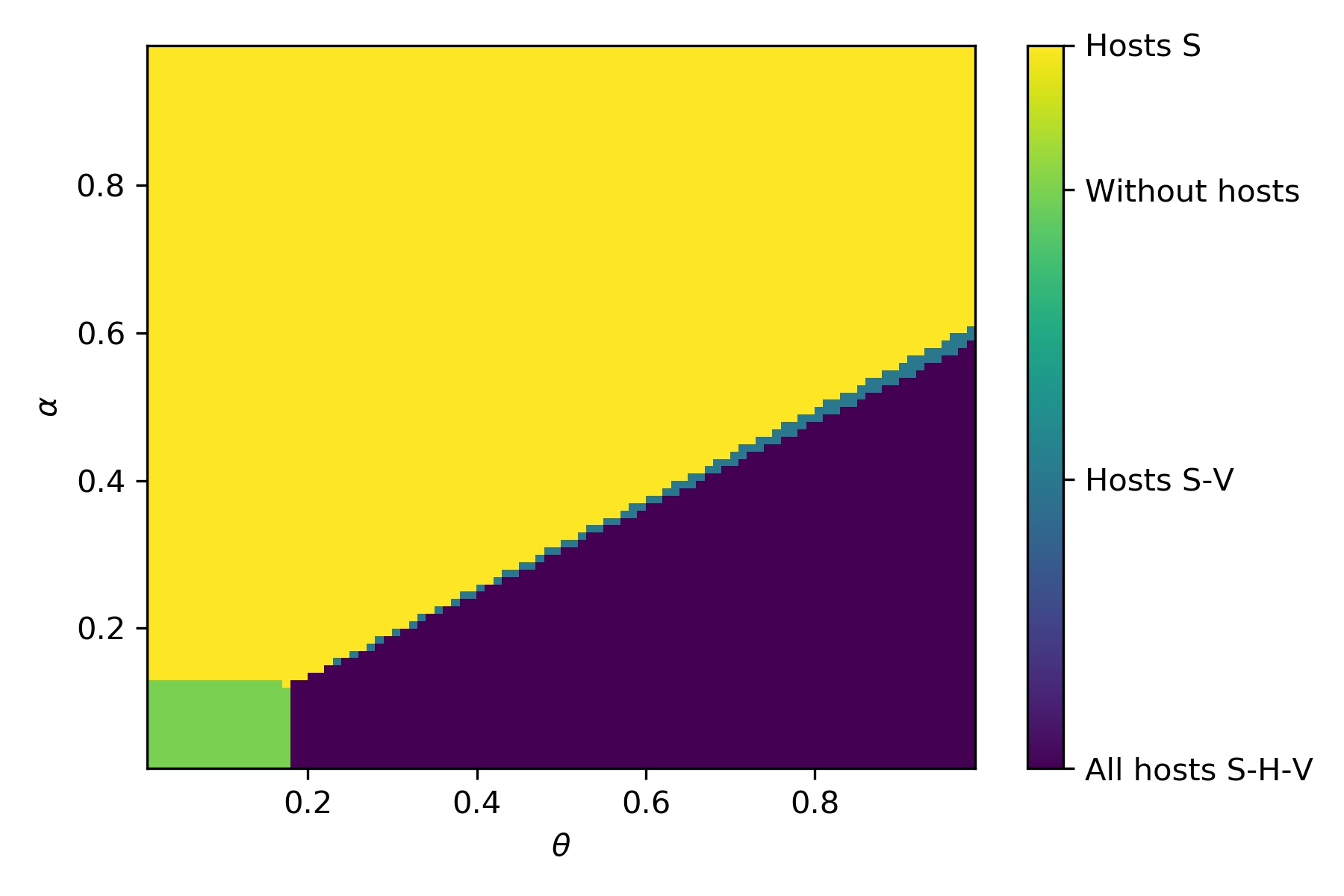}
		
		\caption{ }
	\end{subfigure}
	\begin{subfigure}{.45\textwidth}
		\centering
		\includegraphics[width=1\linewidth]{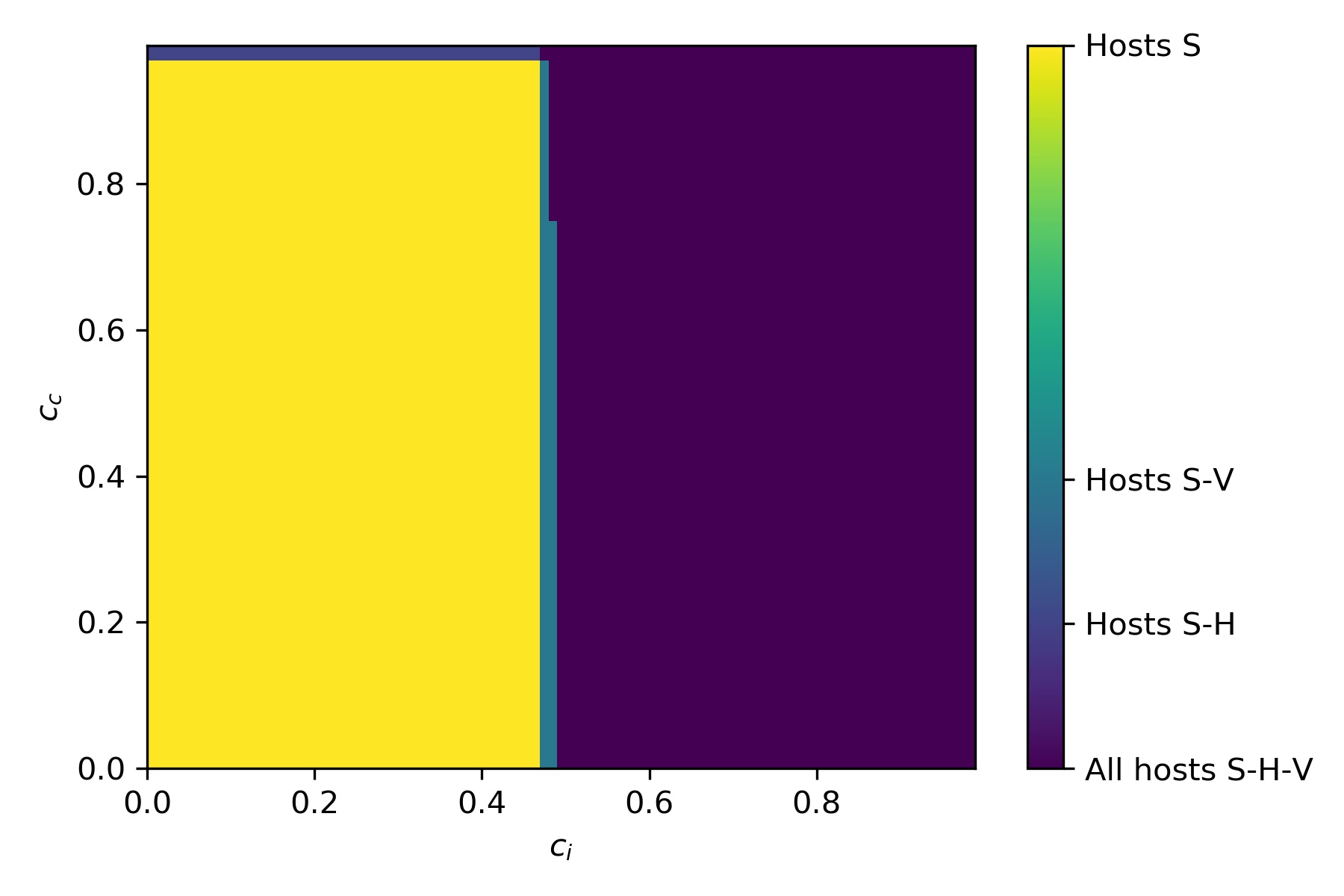}
		
		\caption{ }
	\end{subfigure}
    \caption{a) Population persistence for a range of parasitoid attack survival $\alpha$ and $\theta$ considering the \textit{Medicago sativa} biotype. b) Population persistence for a range of $c_c$ and $c_i$ considering the \textit{Medicago sativa} biotype.}
\end{figure}

\begin{figure}
    \centering
    \begin{subfigure}{.45\textwidth}
		\centering
		\includegraphics[width=1\linewidth]{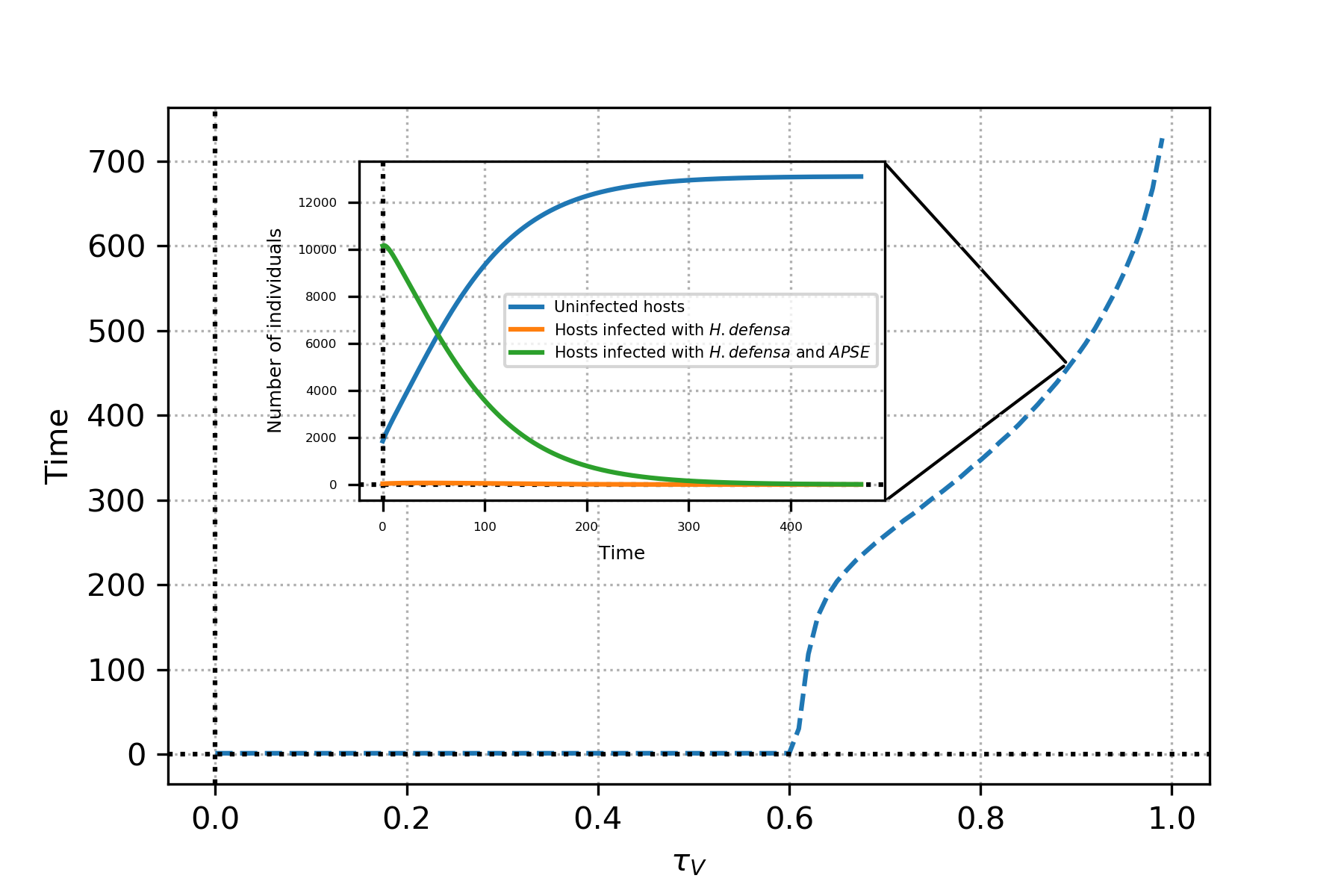}
		
		\caption{}
	\end{subfigure}
	\begin{subfigure}{.45\textwidth}
		\centering
		\includegraphics[width=1\linewidth]{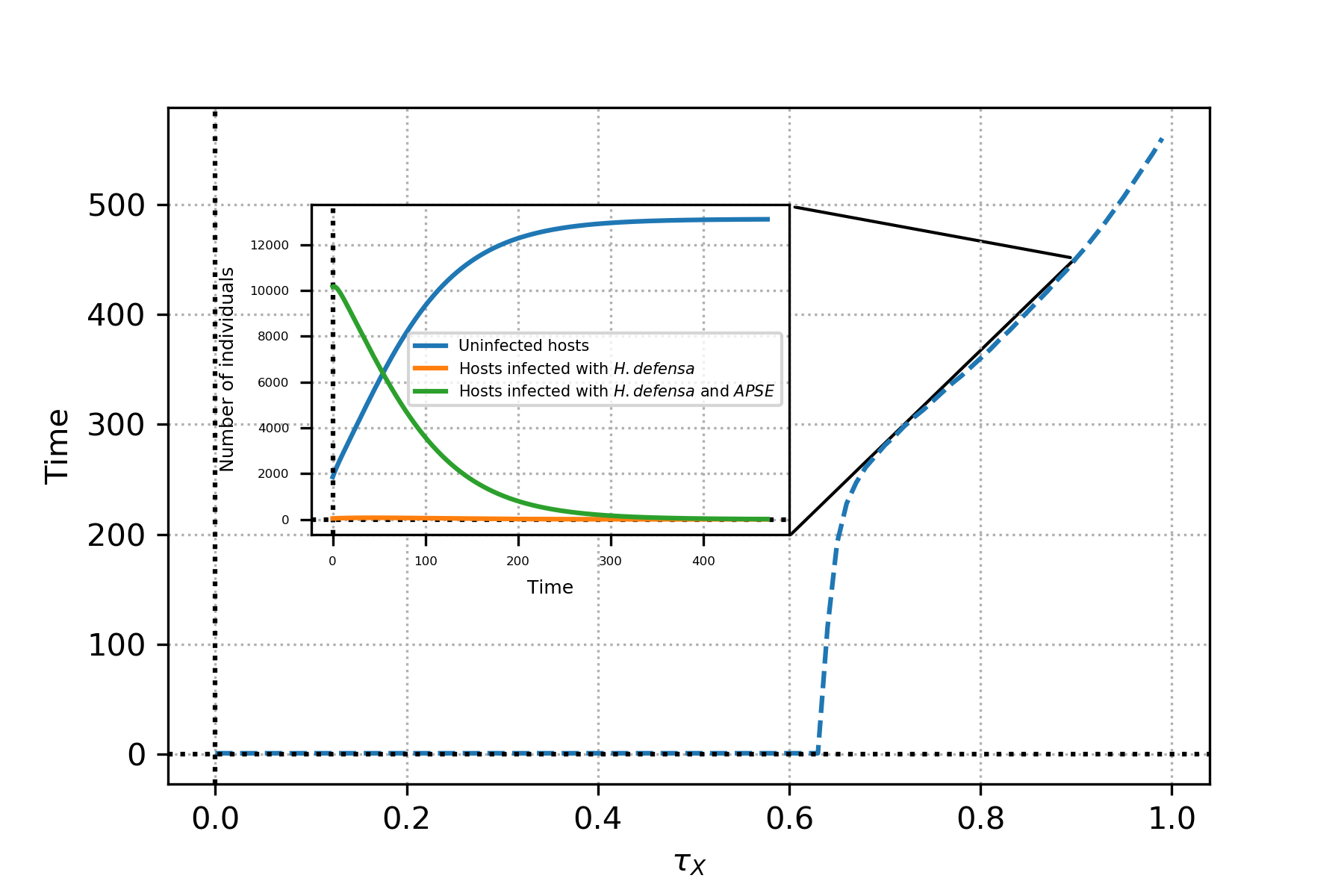}
	
		\caption{}
	\end{subfigure}
    \caption{Time until the population loses resistance as a function of $\tau_V$ and $\tau_X$ considering the biotype \textit{M. sativa}. When the number of iterations reaches $0$ it means that the stability with the presence of parasitoid presented $V=0$.}
    \label{fig:time_loss}
\end{figure}

\section{Discussion}

The influence of microbial symbionts on animal protection against natural enemies is an important open question in ecology \citep{kwiatkowski2012,Vorburger2017, Leybourne2020}. We introduced a new mathematical model based on the bacteriophage effect on parasitism resistance. We have shown that different combinations of host survival after the parasitoid attack for unprotected hosts $S$ plus $H$ ($\alpha$) and protected hosts $V$ ($\theta$) resulted in four possible outcomes: coexistence of all hosts ($S$, $H$ and $V$), extinction of all hosts, the unique presence of uninfected hosts $S$ and infected hosts with \textit{H. defensa} plus \textit{APSE} $V$. The analysis of biotype \textit{G. tinctoria} also illustrated these results, given that our model's unique presence of infected hosts $S$ was a common outcome. This outcome indicates that \textit{H. defensa} can be excluded of \textit{G. tinctoria} biotype population. This finding expanded the results provided by \cite{sochard_data_2020} given that our model provides additional support for the fact that \textit{G. tinctoria} which received \textit{H. defensa} strains of \textit{Medicago sativa} biotype provides no protection against parasitoids. 

Besides, this result also shows the effect of the costs involved in harbouring endosymbionts. We added the constitutive cost $\left(c_c=\frac{b_s-b_h}{b_s}\right)$ and the induced cost $\left(c_i=\frac{b_s-b_v}{b_s}\right)$ considering $b_s \geq b_v \geq b_h$ (i.e Growth rates). Figure 8 indicates that the coexistence of all populations also mainly depends on the induced cost (See figure 8), which supports previous results provided by \cite{kwiatkowski2012}, where they identified employing a mathematical model that the induced cost as a potential driver of coexistence. \cite{kwiatkowski2012} also indicates that the vertical transmission of \textit{H. defensa} can be a potential driver of coexistence, and here we substantiated this finding and complemented their results by showing that the vertical transmission of \textit{APSE}, \textit{H. defensa} by uninfected ($S$) and infected hosts ($V$), hosts' survival after parasitoid attack for unprotected hosts $S$ plus $H$ ($\alpha$) and protected hosts $V$ ($\theta$) can be potential drivers of coexistence.

The triggering mechanisms of the trophic relationships above mentioned have, besides from ecological relevance for trophic systems, a high degree of complexity and are often determinants of success or failure between parasitoid controlling hosts, with significant importance for the biological control of agricultural pests, scenarios especially relevant when the participating microorganisms can confer resistance to the hosts against parasitoids \cite{Vorburger2018}. The use of natural enemies for biological control and integrated pest management in agriculture has been consolidated over time, mainly through the use of microorganisms, such as entomopathogenic fungi and macro-control agents, such as predatory arthropods and mainly parasitoids \cite{Lenteren2018}. In the specific case of parasitoids, the results have been quite promising, especially in the control of pests of critical crops \cite{Chailleux2012, Veiga2013, Lenteren2018}. 

Our results provided by the protection evolution analysis are also applied to the biological control context. Our findings suggest that the vertical transmission of \textit{APSE} and \textit{H. defensa} by $V$ can be potential drivers of the time to the extinction of infected hosts $V$. We observed that these parameters are proportional to when $V$ reaches extinction in a scenario with parasitoids' absence and a high density of protected hosts $V$. This indicates that the host population will become entirely susceptible to parasitoid attack again. Thus, interrupting the release of parasitoids to apply different strategies, such as the release of predators \citep{Parra2019}, has the potential to maintain the high efficiency of biological control with parasitoids. Finally, genetic and genomic techniques are required to observe such details of microbial interference in the host-symbiont-parasitoid system. Also, new standardised methodologies should be developed to optimise the identification of these microorganisms \citep{Vorburger2018} and facilitate the detection for extensive surveys. Finally, as discussed by \cite{Leung2020}, integrating genetics and genomics tend to be the next generation of biological control yielding the procedure optimisation, so our finding is in good agreement with this interdisciplinary field.

To summarise, we have investigated the bacteriophage effect on parasitism resistance by introducing a new mathematical model that considers the microorganism interaction in the host-parasitoid system. We analysed two hosts biotypes by using parameters obtained from literature sources. It allow us to identify that the vertical transmission of \textit{APSE}, \textit{H. defensa} by uninfected ($S$) and infected hosts ($V$), hosts' survival after parasitoid attack for unprotected hosts $S$ plus $H$ ($\alpha$) and protected hosts $V$ ($\theta$) as possible drivers of coexistence. Also, we have shown that vertical transmission of the bacteriophage \textit{APSE} and the vertical transmission of the \textit{H. defensa} by $V$ can be potential drivers of the time to the extinction of infected hosts $V$. Finally, we intend to continue our work to expand the model to consider spatial interactions and add new components such as migration.









\section*{Acknowledgements}
GRP would like to thank Funda\c{c}\~{a}o de Amparo \`{a} Pesquisa do Estado de S\~{a}o Paulo (proc. no. 2014/16609-7 and proc. no. 17/19984-1) for financial support.

\bibliographystyle{plainnat}
\bibliography{refs}

\begin{thebibliography}{39}
\providecommand{\natexlab}[1]{#1}
\providecommand{\url}[1]{\texttt{#1}}
\expandafter\ifx\csname urlstyle\endcsname\relax
  \providecommand{\doi}[1]{doi: #1}\else
  \providecommand{\doi}{doi: \begingroup \urlstyle{rm}\Url}\fi

\bibitem[Bourtzis and Miller(2008)]{bourtzis2008}
Kostas Bourtzis and Thomas~A Miller.
\newblock \emph{Insect Symbiosis}, volume~3.
\newblock CRC Press, 2008.

\bibitem[Brandt et~al.(2017)Brandt, Chevignon, Oliver, and Strand]{Society2017}
Jayce~W Brandt, Germain Chevignon, Kerry~M Oliver, and Michael~R Strand.
\newblock Culture of an aphid heritable symbiont demonstrates its direct role
  in defence against parasitoids.
\newblock \emph{Proceedings of the Royal Society B: Biological Sciences},
  284\penalty0 (1866):\penalty0 20171925, 2017.

\bibitem[Brownlie and Johnson(2009)]{brownlie2009}
Jeremy~C Brownlie and Karyn~N Johnson.
\newblock Symbiont-mediated protection in insect hosts.
\newblock \emph{Trends in microbiology}, 17\penalty0 (8):\penalty0 348--354,
  2009.
\newblock \doi{10.1016/j.tim.2009.05.005}.

\bibitem[Chailleux et~al.(2012)Chailleux, Desneux, Seguret, Do~Thi~Khanh,
  Maignet, and Tabone]{Chailleux2012}
Anaïs Chailleux, Nicolas Desneux, Julien Seguret, Hong Do~Thi~Khanh, Pascal
  Maignet, and Elisabeth Tabone.
\newblock Assessing european egg parasitoids as a mean of controlling the
  invasive south american tomato pinworm tuta absoluta.
\newblock \emph{PLOS ONE}, 7\penalty0 (10):\penalty0 1--8, 10 2012.
\newblock \doi{10.1371/journal.pone.0048068}.
\newblock URL \url{https://doi.org/10.1371/journal.pone.0048068}.

\bibitem[Dion et~al.(2011)Dion, Z{\'{e}}l{\'{e}}, Simon, and
  Outreman]{Dion2011}
E.~Dion, F.~Z{\'{e}}l{\'{e}}, J.~C. Simon, and Y.~Outreman.
\newblock {Rapid evolution of parasitoids when faced with the symbiont-mediated
  resistance of their hosts}.
\newblock \emph{Journal of Evolutionary Biology}, 24\penalty0 (4):\penalty0
  741--750, 2011.
\newblock ISSN 1010061X.
\newblock \doi{10.1111/j.1420-9101.2010.02207.x}.

\bibitem[Ferrari and Vavre(2011)]{ferrari2011}
Julia Ferrari and Fabrice Vavre.
\newblock Bacterial symbionts in insects or the story of communities affecting
  communities.
\newblock \emph{Philosophical Transactions of the Royal Society B: Biological
  Sciences}, 366\penalty0 (1569):\penalty0 1389--1400, 2011.
\newblock \doi{10.1098/rstb.2010.0226}.

\bibitem[Ferreira and Godoy(2014)]{Ferreira2014}
Claudia Ferreira and Wesley Godoy.
\newblock \emph{Ecological Modelling Applied to Entomology}.
\newblock Springer, 01 2014.
\newblock ISBN 978-3-319-06876-3.
\newblock \doi{10.1007/978-3-319-06877-0}.

\bibitem[Frago et~al.(2017)Frago, Mala, Weldegergis, Yang, McLean, Godfray,
  Gols, and Dicke]{Frago2017}
Enric Frago, Mgbrta Mala, Berhane~T. Weldegergis, Chenjiao Yang, Ailsa McLean,
  H.~Charles~J. Godfray, Rieta Gols, and Marcel Dicke.
\newblock {Symbionts protect aphids from parasitic wasps by attenuating
  herbivore-induced plant volatiles}.
\newblock \emph{Nature Communications}, 8\penalty0 (1):\penalty0 1--9, 2017.
\newblock ISSN 20411723.
\newblock \doi{10.1038/s41467-017-01935-0}.
\newblock URL \url{http://dx.doi.org/10.1038/s41467-017-01935-0}.

\bibitem[Gehrer and Vorburger(2012)]{Gehrer2012}
Lukas Gehrer and Christoph Vorburger.
\newblock Parasitoids as vectors of facultative bacterial endosymbionts in
  aphids.
\newblock \emph{Biology letters}, 8\penalty0 (4):\penalty0 613--615, 2012.

\bibitem[Gil and Latorre(2019)]{Gil2019}
Rosario Gil and Amparo Latorre.
\newblock {Unity makes strength: A review on mutualistic symbiosis in
  representative insect clades}.
\newblock \emph{Life}, 9\penalty0 (1):\penalty0 1--24, 2019.
\newblock ISSN 20751729.
\newblock \doi{10.3390/life9010021}.

\bibitem[Giunti et~al.(2015)Giunti, Canale, Messing, Donati, Stefanini,
  Michaud, and Benelli]{Giunti2015}
G.~Giunti, A.~Canale, R.H. Messing, E.~Donati, C.~Stefanini, J.P. Michaud, and
  G.~Benelli.
\newblock Parasitoid learning: Current knowledge and implications for
  biological control.
\newblock \emph{Biological Control}, 90:\penalty0 208--219, 2015.
\newblock ISSN 1049-9644.
\newblock \doi{https://doi.org/10.1016/j.biocontrol.2015.06.007}.
\newblock URL
  \url{https://www.sciencedirect.com/science/article/pii/S1049964415001395}.

\bibitem[Godfray and Godfray(1994)]{Godfray1994}
H~Charles~J Godfray and HCJ Godfray.
\newblock \emph{Parasitoids: behavioral and evolutionary ecology}, volume~67.
\newblock Princeton University Press, 1994.

\bibitem[Haine(2008)]{Haine2008}
Eleanor~R. Haine.
\newblock {Symbiont-mediated protection}.
\newblock \emph{Proceedings of the Royal Society B: Biological Sciences},
  275\penalty0 (1633):\penalty0 353--361, 2008.
\newblock ISSN 14712970.
\newblock \doi{10.1098/rspb.2007.1211}.

\bibitem[Hosokawa and Fukatsu(2020)]{Hosokawa2020}
Takahiro Hosokawa and Takema Fukatsu.
\newblock {Relevance of microbial symbiosis to insect behavior}.
\newblock \emph{Current Opinion in Insect Science}, 39:\penalty0 91--100, 2020.
\newblock ISSN 22145753.
\newblock \doi{10.1016/j.cois.2020.03.004}.
\newblock URL \url{https://doi.org/10.1016/j.cois.2020.03.004}.

\bibitem[Hutchison and Hogg(1985)]{Hutchison1985}
William~D. Hutchison and David~B. Hogg.
\newblock {Time-specific life tables for the pea aphid, Acyrthosiphon pisum
  (Harris), on alfalfa}.
\newblock \emph{Researches on Population Ecology}, 27\penalty0 (2):\penalty0
  231--253, 1985.
\newblock ISSN 00345466.
\newblock \doi{10.1007/BF02515463}.

\bibitem[Kaech and Vorburger(2021)]{Kaech2021}
Heidi Kaech and Christoph Vorburger.
\newblock Horizontal transmission of the heritable protective endosymbiont
  hamiltonella defensa depends on titre and haplotype.
\newblock \emph{Frontiers in Microbiology}, 11, 1 2021.
\newblock ISSN 1664302X.
\newblock \doi{10.3389/fmicb.2020.628755}.

\bibitem[Kaech et~al.(2021)Kaech, Dennis, and Vorburger]{Kaech2021RNA}
Heidi Kaech, Alice~B. Dennis, and Christoph Vorburger.
\newblock Triple rna-seq characterizes aphid gene expression in response to
  infection with unequally virulent strains of the endosymbiont hamiltonella
  defensa.
\newblock \emph{BMC Genomics}, 22, 12 2021.
\newblock ISSN 14712164.
\newblock \doi{10.1186/s12864-021-07742-8}.

\bibitem[Kaech et~al.(2022)Kaech, Jud, Vorburger, Science, and
  Zürich]{Kaech2022}
Heidi Kaech, Stephanie Jud, Christoph Vorburger, Systems Science, and Eth
  Zürich.
\newblock Similar cost of hamiltonella defensa in experimental and natural
  aphid-endosymbiont associations.
\newblock \emph{Ecology and Evolution}, 12:\penalty0 8551, 2022.
\newblock \doi{10.1002/ece3.8551}.
\newblock URL \url{www.ecolevol.org}.

\bibitem[Kwiatkowski and Vorburger(2012)]{kwiatkowski2012}
Marek Kwiatkowski and Christoph Vorburger.
\newblock Modeling the ecology of symbiont-mediated protection against
  parasites.
\newblock \emph{The American Naturalist}, 179\penalty0 (5):\penalty0 595--605,
  2012.
\newblock \doi{10.1086/665003}.

\bibitem[Leung et~al.(2020)Leung, Ras, Ferguson, Ari{\"{e}}ns, Babendreier,
  Bijma, Bourtzis, Brodeur, Bruins, Centuri{\'{o}}n, Chattington,
  Chinchilla-Ram{\'{i}}rez, Dicke, Fatouros, Gonz{\'{a}}lez-Cabrera, Groot,
  Haye, Knapp, Koskinioti, {Le Hesran}, Lyrakis, Paspati, P{\'{e}}rez-Hedo,
  Plouvier, Schl{\"{o}}tterer, Stahl, Thiel, Urbaneja, van~de Zande, Verhulst,
  Vet, Visser, Werren, Xia, Zwaan, Magalh{\~{a}}es, Beukeboom, and
  Pannebakker]{Leung2020}
Kelley Leung, Erica Ras, Kim~B. Ferguson, Simone Ari{\"{e}}ns, Dirk
  Babendreier, Piter Bijma, Kostas Bourtzis, Jacques Brodeur, Margreet~A.
  Bruins, Alejandra Centuri{\'{o}}n, Sophie~R. Chattington, Milena
  Chinchilla-Ram{\'{i}}rez, Marcel Dicke, Nina~E. Fatouros, Joel
  Gonz{\'{a}}lez-Cabrera, Thomas~V.M. Groot, Tim Haye, Markus Knapp, Panagiota
  Koskinioti, Sophie {Le Hesran}, Manolis Lyrakis, Angeliki Paspati, Meritxell
  P{\'{e}}rez-Hedo, Wouter~N. Plouvier, Christian Schl{\"{o}}tterer, Judith~M.
  Stahl, Andra Thiel, Alberto Urbaneja, Louis van~de Zande, Eveline~C.
  Verhulst, Louise~E.M. Vet, Sander Visser, John~H. Werren, Shuwen Xia, Bas~J.
  Zwaan, Sara Magalh{\~{a}}es, Leo~W. Beukeboom, and Bart~A. Pannebakker.
\newblock {Next-generation biological control: the need for integrating
  genetics and genomics}.
\newblock \emph{Biological Reviews}, 95\penalty0 (6):\penalty0 1838--1854,
  2020.
\newblock ISSN 1469185X.
\newblock \doi{10.1111/brv.12641}.

\bibitem[Leybourne et~al.(2020)Leybourne, Bos, Valentine, and
  Karley]{Leybourne2020}
Daniel~J. Leybourne, Jorunn~I.B. Bos, Tracy~A. Valentine, and Alison~J. Karley.
\newblock {The price of protection: a defensive endosymbiont impairs nymph
  growth in the bird cherry-oat aphid, Rhopalosiphum padi}.
\newblock \emph{Insect Science}, 27\penalty0 (1):\penalty0 69--85, 2020.
\newblock ISSN 17447917.
\newblock \doi{10.1111/1744-7917.12606}.

\bibitem[Lu and Kuo(2008)]{Lu2008}
Wei~Nung Lu and Mei~Hwa Kuo.
\newblock {Life table and heat tolerance of Acyrthosiphon pisum (Hemiptera:
  Aphididae) in subtropical Taiwan}.
\newblock \emph{Entomological Science}, 11\penalty0 (3):\penalty0 273--279,
  2008.
\newblock ISSN 13438786.
\newblock \doi{10.1111/j.1479-8298.2008.00274.x}.

\bibitem[Martin and Schwab(2012)]{Martin2012}
Bradford~D. Martin and Ernest Schwab.
\newblock Current usage of symbiosis and associated terminology.
\newblock \emph{International Journal of Biology}, 5, 11 2012.
\newblock ISSN 1916-9671.
\newblock \doi{10.5539/ijb.v5n1p32}.

\bibitem[Moran(2006)]{Moran2006}
Nancy~A. Moran.
\newblock {Symbiosis}.
\newblock \emph{Current Biology}, 16\penalty0 (20):\penalty0 866--871, 2006.
\newblock ISSN 09609822.
\newblock \doi{10.1016/j.cub.2006.09.019}.

\bibitem[Oliver and Higashi(2019)]{Oliver2019}
Kerry~M. Oliver and Clesson~HV Higashi.
\newblock {Variations on a protective theme: Hamiltonella defensa infections in
  aphids variably impact parasitoid success}.
\newblock \emph{Current Opinion in Insect Science}, 32\penalty0
  (March):\penalty0 1--7, 2019.
\newblock ISSN 22145753.
\newblock \doi{10.1016/j.cois.2018.08.009}.
\newblock URL \url{https://doi.org/10.1016/j.cois.2018.08.009}.

\bibitem[Oliver et~al.(2009)Oliver, Degnan, Hunter, and Moran]{oliver2009}
Kerry~M Oliver, Patrick~H Degnan, Martha~S Hunter, and Nancy~A Moran.
\newblock Bacteriophages encode factors required for protection in a symbiotic
  mutualism.
\newblock \emph{Science}, 325\penalty0 (5943):\penalty0 992--994, 2009.
\newblock \doi{10.1126/science.1174463}.

\bibitem[Parra and Coelho(2019)]{Parra2019}
José~Roberto Parra and Junior Coelho, Aloisio.
\newblock {Applied Biological Control in Brazil: From Laboratory Assays to
  Field Application}.
\newblock \emph{Journal of Insect Science}, 19\penalty0 (2), 03 2019.
\newblock ISSN 1536-2442.
\newblock \doi{10.1093/jisesa/iey112}.
\newblock URL \url{https://doi.org/10.1093/jisesa/iey112}.
\newblock 5.

\bibitem[Rajarajan et~al.(2011)Rajarajan, Ibrahim, and Pandian]{Rajarajan2011}
Senguttuvan Rajarajan, Kalibulla~Syed Ibrahim, and Shunmugiah~Karutha Pandian.
\newblock {AP-APSE dpol intein: A novel family A DNA polymerase intein domain.}
\newblock \emph{Bioinformation}, 6\penalty0 (4):\penalty0 149--152, 2011.
\newblock ISSN 09738894.
\newblock \doi{10.6026/97320630006149}.

\bibitem[Rocha et~al.(2018)Rocha, Peterson, Bodin, and Levin]{Rocha2018}
Juan~C. Rocha, Garry Peterson, {\"{O}}rjan Bodin, and Simon~A. Levin.
\newblock {Cascading regime shifts within and across scales}.
\newblock \emph{bioRxiv}, 1383\penalty0 (December):\penalty0 1379--1383, 2018.
\newblock \doi{10.1101/364620}.

\bibitem[Russell(2019)]{Russell2019}
Shelbi~L Russell.
\newblock {Transmission mode is associated with environment type and taxa
  across bacteria-eukaryote symbioses: a systematic review and meta-analysis}.
\newblock \emph{FEMS Microbiology Letters}, 366\penalty0 (3), 01 2019.
\newblock ISSN 0378-1097.
\newblock \doi{10.1093/femsle/fnz013}.
\newblock URL \url{https://doi.org/10.1093/femsle/fnz013}.
\newblock fnz013.

\bibitem[Sochard et~al.(2020{\natexlab{a}})Sochard, Bellec, Simon, and
  Outreman]{Sochard2020}
Corentin Sochard, Laura Bellec, Jean-Christophe Simon, and Yannick Outreman.
\newblock {Influence of “protective” symbionts throughout the different
  steps of an aphid–parasitoid interaction}.
\newblock \emph{Current Zoology}, pages 1--24, 2020{\natexlab{a}}.
\newblock ISSN 1674-5507.
\newblock \doi{10.1093/cz/zoaa053}.

\bibitem[Sochard et~al.(2020{\natexlab{b}})Sochard, Bellec, Simon, and
  Outreman]{sochard_data_2020}
Corentin Sochard, Laura Bellec, Jean-Christophe Simon, and Yannick Outreman.
\newblock {Influence of 'protective' symbionts throughout the different steps
  of an aphid-parasitoid interaction}.
\newblock \emph{Current Zoology}, September 2020{\natexlab{b}}.
\newblock \doi{10.1093/cz/zoaa053}.
\newblock URL \url{https://doi.org/10.1093/cz/zoaa053}.
\newblock Funding: ANR Hmicmac 16-CE02-0014.

\bibitem[{Van Der Wilk} et~al.(1999){Van Der Wilk}, Dullemans, Verbeek, and
  {Van Den Heuvel}]{VanDerWilk1999}
Frank {Van Der Wilk}, Annette~M. Dullemans, Martin Verbeek, and Johannes~F.J.M.
  {Van Den Heuvel}.
\newblock {Isolation and characterization of APSE-1, a bacteriophage infecting
  the secondary endosymbiont of Acyrthosiphon pisum}.
\newblock \emph{Virology}, 262\penalty0 (1):\penalty0 104--113, 1999.
\newblock ISSN 00426822.
\newblock \doi{10.1006/viro.1999.9902}.

\bibitem[van Lenteren et~al.(2018)van Lenteren, Bolckmans, K{\"o}hl,
  Ravensberg, and Urbaneja]{Lenteren2018}
Joop~C van Lenteren, Karel Bolckmans, J{\"u}rgen K{\"o}hl, Willem~J Ravensberg,
  and Alberto Urbaneja.
\newblock Biological control using invertebrates and microorganisms: plenty of
  new opportunities.
\newblock \emph{BioControl}, 63\penalty0 (1):\penalty0 39--59, 2018.

\bibitem[Veiga et~al.(2013)Veiga, Vacari, Volpe, de~Laurentis, and
  Bortoli]{Veiga2013}
Ana~C.P. Veiga, Alessandra~M. Vacari, Haroldo~X.L. Volpe, Valéria~L.
  de~Laurentis, and Sergio A.~De Bortoli.
\newblock Quality control of cotesia flavipes (cameron) (hymenoptera:
  Braconidae) from different brazilian bio-factories.
\newblock \emph{Biocontrol Science and Technology}, 23\penalty0 (6):\penalty0
  665--673, 2013.
\newblock \doi{10.1080/09583157.2013.790932}.

\bibitem[Vorburger(2018)]{Vorburger2018}
Christoph Vorburger.
\newblock {Symbiont-conferred resistance to parasitoids in aphids –
  Challenges for biological control}.
\newblock \emph{Biological Control}, 116\penalty0 (February):\penalty0 17--26,
  2018.
\newblock ISSN 10499644.
\newblock \doi{10.1016/j.biocontrol.2017.02.004}.
\newblock URL \url{http://dx.doi.org/10.1016/j.biocontrol.2017.02.004}.

\bibitem[Vorburger et~al.(2013)Vorburger, Ganesanandamoorthy, and
  Kwiatkowski]{vorburger2013}
Christoph Vorburger, Pravin Ganesanandamoorthy, and Marek Kwiatkowski.
\newblock Comparing constitutive and induced costs of symbiont-conferred
  resistance to parasitoids in aphids.
\newblock \emph{Ecology and evolution}, 3\penalty0 (3):\penalty0 706--713,
  2013.

\bibitem[Vorburger et~al.(2017)Vorburger, Siegrist, and Rhyner]{Vorburger2017}
Christoph Vorburger, Gabrielle Siegrist, and Nicola Rhyner.
\newblock {Faithful vertical transmission but ineffective horizontal
  transmission of bacterial endosymbionts during sexual reproduction of the
  black bean aphid, Aphis fabae}.
\newblock \emph{Ecological Entomology}, 42\penalty0 (2):\penalty0 202--209,
  2017.
\newblock ISSN 13652311.
\newblock \doi{10.1111/een.12379}.

\bibitem[Weldon et~al.(2013)Weldon, Strand, and Oliver]{weldon2013phage}
SR~Weldon, MR~Strand, and KM~Oliver.
\newblock Phage loss and the breakdown of a defensive symbiosis in aphids.
\newblock \emph{Proceedings of the Royal Society B: Biological Sciences},
  280\penalty0 (1751):\penalty0 20122103, 2013.

\end{thebibliography}

\end{document}